\documentclass[twocolumn,,twocolappendix,trackchanges]{aastex701}

\usepackage[T1]{fontenc}
\usepackage{siunitx} 
\usepackage{multirow}
\usepackage{booktabs}
\usepackage{comment}
\usepackage{CJK}

\shorttitle{He WD and CO WD in Partial TDEs}
\shortauthors{Chong et al.}

\begin{document}
\begin{CJK*}{UTF8}{gbsn}

\title{Stripped White Dwarfs as Spectroscopic Clocks: Abundance Transitions in Partial Tidal Disruptions}

%% Author and Affiliation (v7.0.1 requires emails for all authors)
\author[0009-0008-0168-0032]{Taylor Yi-Ching Chong}
\email{u3630897@connect.hku.hk}
\affiliation{Department of Physics, University of Hong Kong, Pokfulam Road, Hong Kong, People's Republic of China}
\affiliation{The Hong Kong Institute for Astronomy and Astrophysics, University of Hong Kong, Hong Kong, China}

\author[0000-0002-3525-791X]{Jin-Hong Chen (陈劲鸿)}
\email{chenjh2@hku.hk}
\correspondingauthor{Jin-Hong Chen (陈劲鸿)}
\email{chenjh2@hku.hk}
\affiliation{Department of Physics, University of Hong Kong, Pokfulam Road, Hong Kong, People's Republic of China}
\affiliation{The Hong Kong Institute for Astronomy and Astrophysics, University of Hong Kong, Hong Kong, China}
\affiliation{Shenzhen Institute of Research and Innovation, The University of Hong Kong, Shenzhen 518057, P. R. China}

\author[0000-0002-9589-5235]{Lixin Dai}
\email{lixindai@hku.hk}
\affiliation{Department of Physics, University of Hong Kong, Pokfulam Road, Hong Kong, People's Republic of China}
\affiliation{The Hong Kong Institute for Astronomy and Astrophysics, University of Hong Kong, Hong Kong, China}

%% Abstract
\begin{abstract}
We investigate partial tidal disruptions of He and CO white dwarfs (WDs) by a $10^{4}\,M_{\odot}$ intermediate-mass black hole (IMBH) using three-dimensional hydrodynamic simulations. We construct realistic WD structures and abundance profiles with \texttt{MESA} and map them into \texttt{FLASH} to follow the remnant response and the time-dependent composition of the fallback debris in single pericentre passages on tightly bound orbits ($e=0.9$). The element-specific fallback rates exhibit pre-peak compositional transitions that trace the WD's stratified layers. In the grazing encounter ($\beta=0.55$), hydrogen dominates the earliest returning debris, followed by helium, showing a H$\rightarrow$He transition. Deeper encounters ($\beta=0.7$ and $0.8$) penetrate the CO core and produce a He$\rightarrow$C transition, that occurs earlier relative to peak fallback and is stronger in the deeper encounter. These signatures provide a potential spectroscopic diagnostic for identifying CO-WD tidal disruption events (TDEs). We then develop a semi-analytic model for repeated tidal stripping and predict passage-to-passage compositional evolution, in which the stripped material becomes progressively enriched in heavier elements as deeper layers are exposed, leading to stronger and earlier abundance transitions in later flares. We find that WDs can survive several dozen ($\sim 30$--$80$) passages, depending on their initial mass and compactness. Our results link time-dependent abundance signatures to repeating WD disruptions and provide testable observable predictions for identifying IMBH-powered transients. 
\end{abstract}

%% Keywords: Using AAS standard identifiers

\section{Introduction} \label{sec:intro}

When a star approaches too close to a massive black hole (MBH), it could be torn apart by the tidal force of MBH. In typical scenarios, half of the stellar material will be bound and then accreted onto the MBH, while another half will be ejected with very high speed to hyperbolic trajectories \citep{1975Natur.254..295H, 1978MNRAS.184...87F, 1979Natur.280..214G, 1982Natur.296..211C, 1988Natur.333..523R}. These phenomena usually give rise to bright flares in the sky, known as tidal disruption events (TDEs). White dwarfs (WDs) can only be disrupted by intermediate-mass black holes (IMBHs) with masses below $\sim10^{5}\,M_{\odot}$, providing a unique electromagnetic and gravitational-wave probe of otherwise elusive IMBH populations \citep{2014ApJ...794....9M, 2023ApJ...947...32C}. Understanding WD-IMBH interactions is therefore important for constraining the population of IMBHs and their possible role as seeds of supermassive black holes \citep{2017IJMPD..2630021M}.

Unlike the commonly studied parabolic TDEs, WD-IMBH encounters can occur on tightly bound eccentric orbits. In such systems, a WD may survive the first encounter and return for subsequent passages, experiencing repeated partial tidal stripping rather than complete disruption \citep{2014ApJ...794....9M,2023ApJ...948...89K, 2023ApJ...944..184L}. Most of the observed quasi-periodic eruptions (QPEs) have been attributed to tidal stripping on highly eccentric orbits with $e > 0.9$ \citep{2022ApJ...930..122C}, including sources such as GSN 069, eRO-QPE1 and eRO-QPE2 \citep{2020MNRAS.493L.120K, 2022ApJ...933..225W, 2022A&A...661A..55Z}. The repeated stripping of a WD provides a potential opportunity to investigate both the orbital evolution of compact objects and the internal structure of the disrupted star.

The fallback rate of stellar debris is a key quantity connecting the disruption properties to the resulting light curve \citep{2019ApJ...872..151M}. Previous hydrodynamic studies have demonstrated that the fallback rate depends strongly on stellar structure and orbital parameters \citep{2009ApJ...697L..77R, 2013ApJ...767...25G}. However, most numerical studies of TDEs have focused on single encounters, and have adopted simplified polytropic stellar models. The long-term evolution of WD-IMBH systems undergoing multiple tidal encounters has been studied using analytical methods \citep{2013ApJ...777..133M, 2014ApJ...794....9M, 2023ApJ...944..184L} or semi-analytical model accounts for mass loss and tidal excitation in \citet{2024OJAp....7E..48B}, which have provided valuable estimates of the mass loss per passage and the resulting long-term orbital evolution. However, the time-dependent compositional evolution of the stripped material over the course of repeated encounters has not yet been examined in detail, since these studies did not track how the star's stratified internal structure is progressively removed and imprinted on the fallback debris.

Realistic stellar structures are especially important for WDs because their density and composition profiles are strongly stratified, and this stratification can imprint itself on the debris energy distribution and thus on the fallback rate \citep{2020ApJ...905..141L}. For main-sequence stars, \citet{2019ApJ...882L..25L} used \texttt{MESA} models to track the composition of fallback material as a function of time and found substantial differences relative to both polytropic hydrodynamical simulations \citep{2013ApJ...767...25G} and analytical estimates of compositional fallback \citep{2018ApJ...857..109G}. Similarly, \citet{2019ApJ...882L..26G} showed that using realistic \texttt{MESA} stellar models (rather than polytropes) can change inferred black hole masses by up to an order of magnitude. By contrast, analogous studies for WDs, especially in partial and repeating encounters, where time-dependent abundance transitions may provide a distinctive diagnostic, remain comparatively unexplored. Among different types of WDs, $\sim 0.6 M_{\odot}$ CO WDs are the most common one, which is at the peak of the WD mass distribution \citep{2016MNRAS.461.2100T, 2020MNRAS.499.1890M}, making them the most relevant case for WD-IMBH tidal interactions. 

In this work, we simulate partial tidal disruptions of He and CO WDs by an IMBH using realistic \texttt{MESA} stellar models mapped into three-dimensional hydrodynamics. We ask (i) how the WD's layered composition imprints time-dependent, element-specific transitions in the fallback debris, potentially enabling spectroscopic diagnostics that distinguish CO-WD TDEs from other disruptions, and (ii) how repeating TDEs progressively expose deeper layers, altering the fallback composition and driving long-term mass loss.

The remainder of this paper first follows a WD through a single tidal encounter and then extrapolates to the sequence of passages in a repeating disruption. Section~\ref{sec:Tidal Stripping of a White Dwarf} describes our \texttt{MESA} WD models, numerical setup, and methods for computing the fallback rate and stripped mass. Section~\ref{sec:Simulation results} presents our simulation results, including the time-dependent elemental abundances of the fallback debris of He and CO WDs in a single pericentre passage. Section~\ref{sec:theory_repeating_tdes} extends these results to multiple passages, quantifying the stripped mass (and its composition) per orbit and estimating the maximum number of survivable encounters. Section~\ref{sec:discussion} discusses observational implications (including possible jet launching) and candidate sources, and Section~\ref{sec:conclusion} summarizes our findings.

\section{Tidal Stripping of a White Dwarf} \label{sec:Tidal Stripping of a White Dwarf}
\subsection{WD Models} \label{sec:WD Models}
We construct WD models with the 1D stellar evolution code \texttt{MESA} \citep{2011ApJS..192....3P} and simulate their tidal disruption using \texttt{FLASH}, a 3D grid-based adaptive mesh refinement hydrodynamics code \citep{2000ApJS..131..273F}. We focus on a $0.15\ M_{\odot}$ He-core WD and a $0.6\ M_{\odot}$ CO-core WD disrupted by a $M_{\rm h} = 10^{4}\ M_{\odot}$ black hole on an eccentric orbit with $e=0.9$. We consider impact parameters $\beta=0.55,\,0.7,\,0.8$ (with $\beta=0.7$ used for the He WD).

To quantify realistic WD structure, we construct seven \texttt{MESA} WD models spanning a range of masses and core compositions, two of which are subsequently mapped into \texttt{FLASH} for the 3D disruption simulations. Each model is evolved from the pre--main sequence through to the WD cooling track. We adopt the initial abundances listed in Table~\ref{tab:MESA_Parameters} and use the reaction networks \texttt{o18\_and\_ne22.net}, \texttt{co\_burn\_extras.net}, and \texttt{co\_burn\_plus.net} to produce He, CO, and ONe WDs, respectively. We track $10$--$16$ elements, but for clarity we present only seven representative species with the largest mass fractions.

\begin{deluxetable}{lccccc}
\tabletypesize{\footnotesize}
\tablecaption{\texttt{MESA} Initial Parameters \label{tab:MESA_Parameters}}
\tablewidth{0pt}
\tablehead{
\colhead{Core Type} & \colhead{Progenitor} & \colhead{WD Mass} & \colhead{X} & \colhead{Y} & \colhead{Z} \\
\colhead{} & \colhead{($M_\odot$)} & \colhead{($M_\odot$)} & \colhead{} & \colhead{} & \colhead{}
}
\startdata
He core & 1.50 & 0.15 & 0.72 & 0.27 & 0.01 \\
        & 2.00 & 0.28 & 0.72 & 0.27 & 0.01 \\
        & 2.25 & 0.32 & 0.72 & 0.27 & 0.01 \\
\tableline
CO core & 3.10 & 0.60 & 0.67 & 0.31 & 0.02 \\
        & 4.50 & 0.86 & 0.67 & 0.31 & 0.02 \\
        & 6.00 & 0.98 & 0.67 & 0.31 & 0.02 \\
\tableline
ONe core & 11.00 & 1.11 & 0.69 & 0.28 & 0.02 \\
\enddata
%\tablecomments{}
\end{deluxetable}

\begin{deluxetable}{lcccc}
\tabletypesize{\footnotesize}
\tablecaption{Stellar Model Parameters \label{tab:stellar_data}}
\tablewidth{0pt}
\tablehead{
\colhead{Profile} & \colhead{Mass} & \colhead{Radius} & \colhead{Central Density} & \colhead{Mean Density} \\
\colhead{} & \colhead{($M_\odot$)} & \colhead{($R_\odot$)} & \colhead{(g cm$^{-3}$)} & \colhead{(g cm$^{-3}$)}
}
\startdata
He01  & 0.15 & 0.0229 & $1.12 \times 10^5$ & $1.15 \times 10^4$ \\
He02  & 0.28 & 0.0178 & $4.47 \times 10^5$ & $4.75 \times 10^4$ \\
He03  & 0.32 & 0.0176 & $6.09 \times 10^5$ & $5.88 \times 10^4$ \\
\tableline
CO01  & 0.60 & 0.0154 & $3.34 \times 10^6$ & $3.36 \times 10^5$ \\
CO02  & 0.86 & 0.0102 & $1.48 \times 10^7$ & $1.40 \times 10^6$ \\
CO03  & 0.98 & 0.0085 & $3.02 \times 10^7$ & $2.62 \times 10^6$ \\
\tableline
ONe01 & 1.11 & 0.0067 & $7.57 \times 10^7$ & $6.64 \times 10^6$ \\
\enddata
\end{deluxetable}

\begin{figure}[htbp]
\centering
\plotone{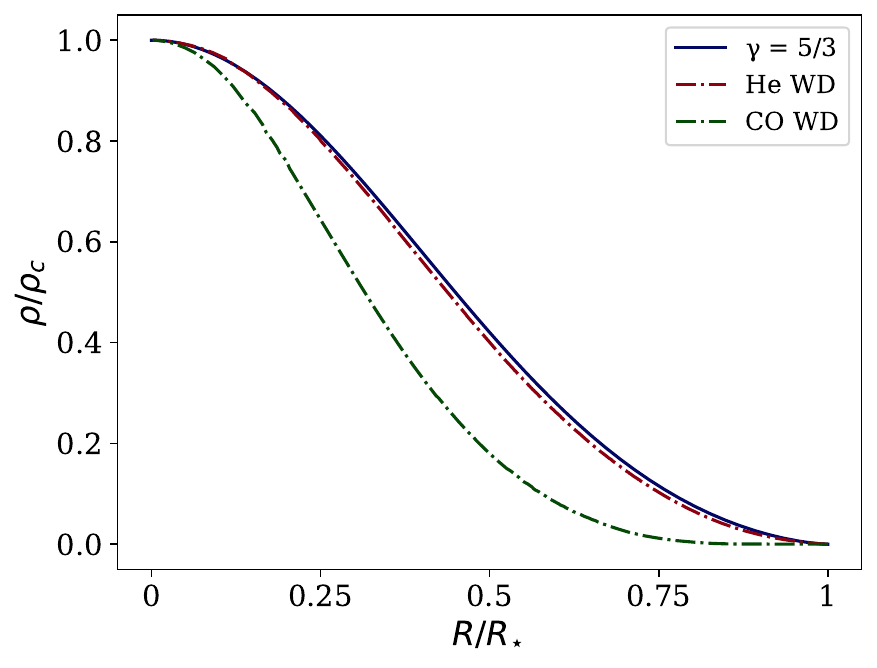}
\caption{Normalized density profiles of a $0.15\,M_{\odot}$ He WD and a $0.6\,M_{\odot}$ CO WD constructed with \texttt{MESA}. The radius is normalized by $R_{\star}$ and the density by the central density. \label{fig:Density_profiles}}
\end{figure}

\begin{figure}[htbp]
\plotone{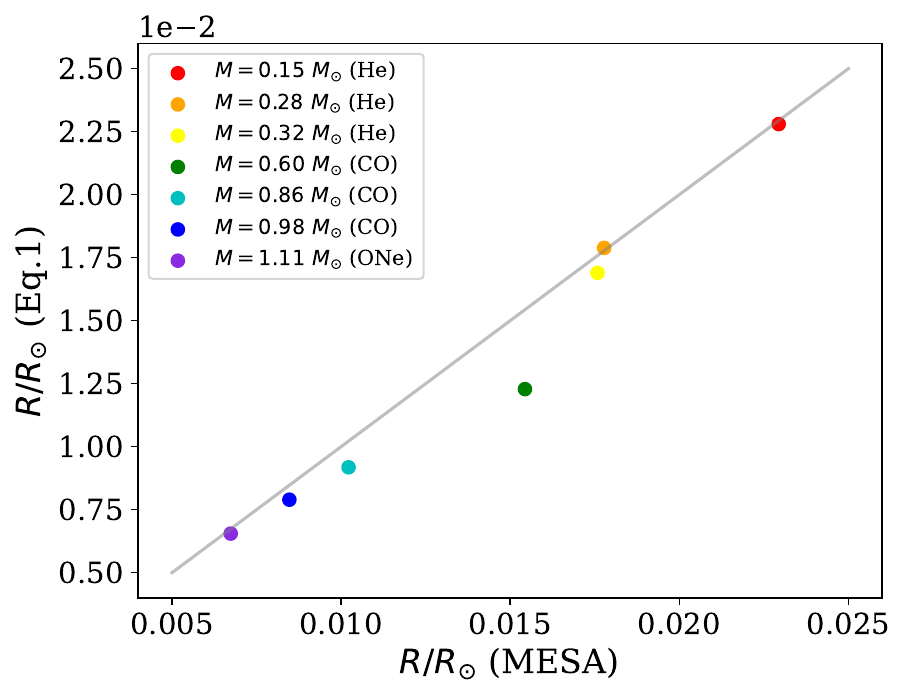}
\caption{Comparison between WD radii from \texttt{MESA} models and those predicted by the analytic mass--radius relation. Points span $0.15$--$1.11\,M_{\odot}$ and core compositions He/CO/ONe. The grey line indicates 1:1 agreement.
\label{fig:mesa_vs_polytrope_radius}}
\end{figure}

Figure~\ref{fig:Density_profiles} shows the normalized density profiles of our $0.15\,M_{\odot}$ He WD and $0.6\,M_{\odot}$ CO WD. The density profile sets the degree of central concentration and therefore influences a WD's susceptibility to tidal disruption \citep{2019ApJ...882L..25L}. The central densities are $\rho_{\mathrm{c}}\approx 10^{5}\,\mathrm{g\,cm^{-3}}$ (He) and $\rho_{\mathrm{c}}\approx 3\times 10^{6}\,\mathrm{g\,cm^{-3}}$ (CO), so the CO WD is denser by a factor of $\sim 30$ and is correspondingly more resistant to stripping. The density profile of the He WD is well approximated by a $\Gamma=5/3$ polytrope, while that of the CO WD is slightly deviated from the  polytrope. This motivates us to use a more realistic WD's density profile to study the tidal disruption of WDs. Figure~\ref{fig:mesa_vs_polytrope_radius} shows that the \texttt{MESA} radii are consistent with the analytic mass--radius relation from \citet{1972ApJ...175..417N}:
\begin{equation}
    R_{\star}(M_{\star}) \simeq 0.022\,\mu_{\mathrm{e}}^{-1} R_{\odot}\left(\frac{M_{\star}}{M_{\mathrm{Ch}}}\right)^{-1/3}
    \left[1 - \left(\frac{M_{\star}}{M_{\mathrm{Ch}}}\right)^{4/3}\right]^{1/2},
    \label{eqt:mass_radius_relationship}
\end{equation}
where $M_{\mathrm{Ch}}\simeq 1.44\,M_{\odot}$ is the Chandrasekhar mass and $\mu_{\mathrm{e}}=2$ is the mean molecular weight per electron.

\begin{figure*}[ht!]
\centering
\gridline{\fig{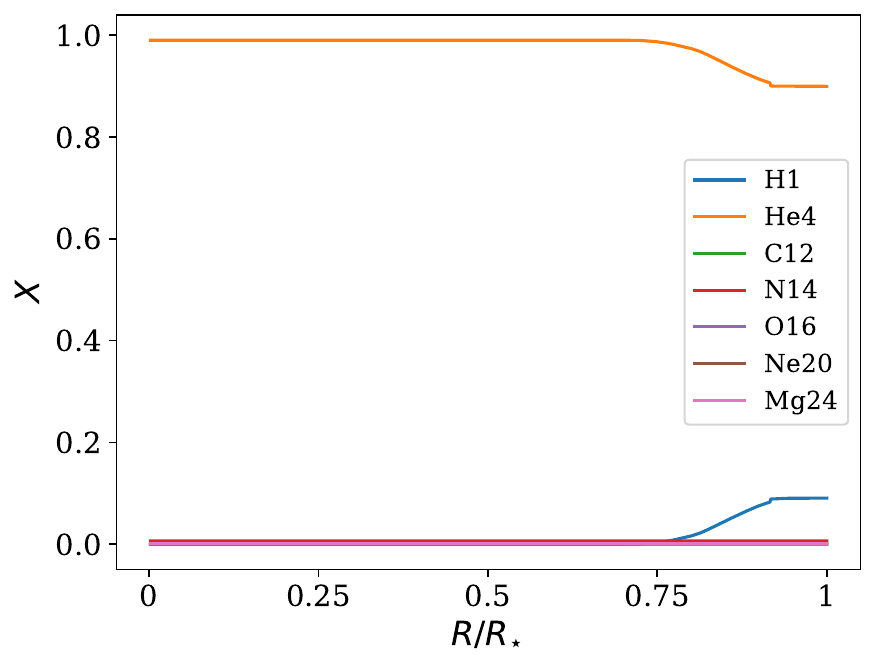}{0.48\textwidth}{(a) He WD abundance profile}
          \fig{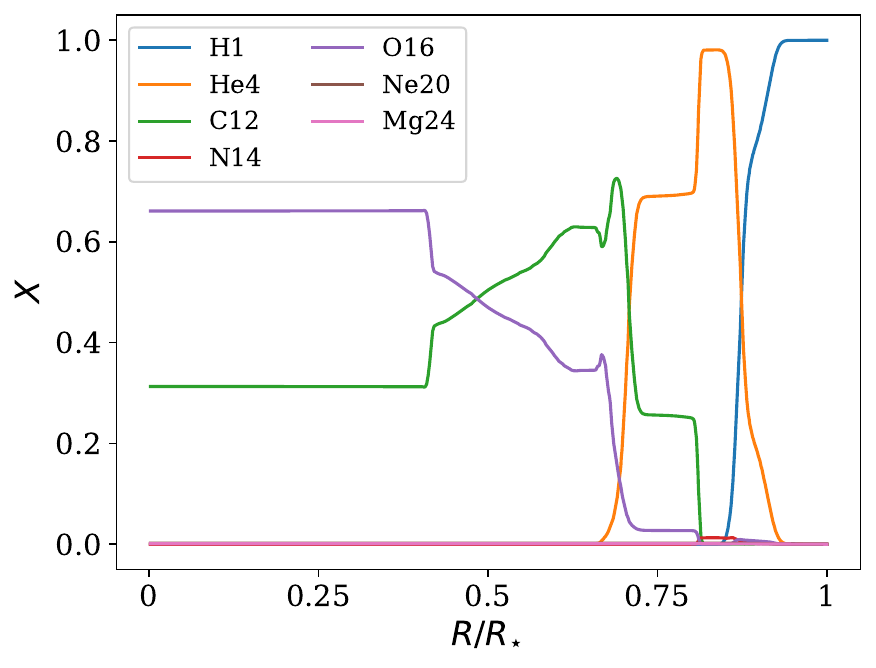}{0.48\textwidth}{(b) CO WD abundance profile}
         }
\caption{Abundance profiles of a $0.15\,M_{\odot}$ He WD and a $0.6\,M_{\odot}$ CO WD constructed with \texttt{MESA}. The radius is normalized by $R_{\star}$. \label{fig:Abundance profile}}
\end{figure*}

\texttt{MESA} also provides detailed abundance profiles. Figure~\ref{fig:Abundance profile} shows that the He WD consists of a He core with a thin H envelope, whereas the CO WD has a CO core surrounded by a He layer and an outer H envelope. Mapping these stratified abundance profiles into the 3D hydrodynamics enables us to track how successive passages peel off distinct compositional layers and imprint them on the fallback debris \citep{2019ApJ...882L..25L}.

\subsection{Hydrodynamics Simulation Setup} \label{sec:Hydrodynamics simulation Setup}
We map the 1D \texttt{MESA} profiles of density, pressure, temperature, and composition for the $0.15\,M_{\odot}$ He WD and $0.6\,M_{\odot}$ CO WD onto a 3D Cartesian grid in \texttt{FLASH}. We perform four simulations with $e=0.9$: one He-WD run with $\beta=0.7$ and three CO-WD runs with $\beta=0.55,\,0.7,\,0.8$. Simulations are carried out in the WD rest frame. We use the directionally split piecewise-parabolic method \citep{1984JCoPh..54..174C} implemented in \texttt{FLASH}, together with the modified gravity algorithm of \citet{2011ApJ...732...74G}; details of our multipole gravity settings follow \citet{2013ApJ...767...25G}. The gas evolves adiabatically with an equation of state parameter $\gamma=5/3$.

The initial separation from the $10^{4}\,M_{\odot}$ IMBH is $5\,R_{\mathrm{T}}$, where $R_{\rm T} = (M_{\rm h} / M_{\star})^{1/3}R_{\star}$ is the tidal radius. We relax the mapped star for $50\,\mathrm{s}$ to obtain numerical stability. The computational domain spans $3\times 10^{12}\,\mathrm{cm}$ on each side and is initialized with $8\times 8\times 8$ blocks, which are adaptively refined based on density. The maximum refinement level is 16 for the $0.15\,M_{\odot}$ He WD and 15 for the $0.6\,M_{\odot}$ CO WD.

Because we simulate a single encounter, we treat the WD orbit as Keplerian and neglect relativistic effects (e.g., apsidal precession and gravitational-wave energy loss), which have small influence for our parameters. We track seven elemental mass fractions in the debris. Our primary diagnostics are the fallback rate and the time-dependent composition of the fallback material.

\subsection{Calculation Methods for Mass Fallback Rate and Stripped Mass} \label{sec:Calculation Methods for Mass Fallback Rate and Stripped Mass}
After pericentre passage, bound debris returns to pericentre in order of binding energy: the most bound material returns first, followed by progressively less bound material. In a partial disruption, part of the stellar mass remains bound to the surviving WD remnant, so identifying the debris bound to the black hole requires separating remnant-bound material from BH-bound material. We compute the stripped mass using the iterative energy-based method described in \citet{2013ApJ...767...25G}.

From the simulations we extract the specific orbital-energy distribution of the stripped debris relative to the BH, $dM/dE$. Combining this with $dE/dt$ from Kepler's third law yields the fallback rate
\begin{equation}
    \dot{M}(t) = \frac{2\pi}{3}\,(G M_{\mathrm{h}})^{2/3}\,\frac{dM}{dE}\,t^{-5/3},
\end{equation}
where we assume $dM/dE$ has ``frozen in'' once the debris has traveled sufficiently far from the BH (we take $r\ge 10\,R_{\mathrm{T}}$), at which point the distribution has converged. In computing $dM/dE$ we exclude material still bound to the WD remnant, which will not accrete onto the BH and instead participates in subsequent encounters \citep{2023ApJ...947...32C}.

\subsection{Analytical Calculation of Fallback Timescale and Stripped Layer Depth} \label{sec:analytical}
In this section, we present analytic estimates for the fallback timescale, the stripped mass per passage, and the corresponding stripped-layer depth.

The return time of a given debris element is set by its specific orbital energy. Following \citet{2009MNRAS.392..332L}, we assume a frozen-in energy distribution spanning $[-\Delta\epsilon,\,+\Delta\epsilon]$, with
$\Delta\epsilon\simeq \xi G M_{\mathrm{h}} R_{\star}/R_{\mathrm{p}}^{2}$. The most bound debris then has
\begin{equation}
    \epsilon_{\mathrm{mb}} \simeq -\frac{G M_{\mathrm{h}}}{2a} - \frac{\xi G M_{\mathrm{h}} R_{\star}}{R_{\mathrm{p}}^{2}},
\end{equation}
where $a=R_{\rm p}/(1-e)$ is the semi-major axis. The factor $\xi$ captures deviations from the simplest impulse estimate; we adopt $\xi=3.5$, following \citet{2023ApJ...947...32C}, for which this value provides a good approximation to the simulated $\Delta\epsilon$. Using Kepler's third law, $P^{2}=(4\pi^{2}/G M_{\mathrm{h}})a^{3}$, we estimate the fallback timescale in terms of the orbital period $P$ as
\begin{equation}
    t_{\mathrm{fb}} = P\left[1 + \frac{2\xi\beta}{(1-e)}\left(\frac{M_{\mathrm{h}}}{M_{\star}}\right)^{-1/3}\right]^{-3/2}.
    \label{eqt:fallback_timescale}
\end{equation}

For eccentric disruptions ($e<1$), both the leading and trailing debris streams are generally bound to the BH and can return to pericentre. The fallback time $t_{\mathrm{fb}}$ corresponds to the return of the leading (most tightly bound) stream; the trailing stream may return later. Remnant survival is therefore essential for repeated stripping over many orbits. We use our single-encounter simulations (Section~\ref{sec:Simulation results}) to characterize the response of He and CO WDs, and in Section~\ref{sec:theory_repeating_tdes} we assess how this response accumulates over repeated passages.

Once mass loss begins, a WD can lose additional mass at each subsequent pericentre passage. We estimate the stripped mass per encounter using the analytic approach of \citet{2010MNRAS.409L..25Z}, together with the mass--radius relation in Equation~\ref{eqt:mass_radius_relationship}. We approximate the stripped layer as a thin spherical shell of overflow depth $\Delta = R_{\star} - R_{\mathrm{roche}}$ above the Roche lobe, so that
\begin{equation}
    \delta M = 4\pi R_{\star}^{2}\int_{0}^{\Delta}\rho(z)\,dz,
    \label{eqt:stripped mass integral}
\end{equation}
where z is measured inward from the WD surface to the Roche Lobe. 

Assuming a polytropic equation of state $P=K\rho^{\Gamma}$ with $\Gamma=5/3$ and integrating the hydrostatic equation near the surface,
\begin{equation}
    \frac{dP}{dz} = \frac{G M_{\star}}{R_{\star}^{2}}\,\rho,
\end{equation}
we obtain the near-surface density profile
\begin{equation}
    \rho(z) = \left(\frac{2G M_{\star}}{5K R_{\star}^{2}}\right)^{3/2} z^{3/2}
           = \left(\frac{2G M_{\star}}{5K R_{\star}}\right)^{3/2}\left(\frac{z}{R_{\star}}\right)^{3/2}.
    \label{eqt:density}
\end{equation}
Carrying out the integral in Equation~\ref{eqt:stripped mass integral} and substituting $R(M_{\star})$ (Equation~\ref{eqt:mass_radius_relationship}) yields
\begin{equation}
    \frac{\delta M}{M_{\star}} = 4.5\left[1 - \left(\frac{M_{\star}}{M_{\mathrm{Ch}}}\right)^{4/3}\right]^{3/4}\left(\frac{\Delta}{R}\right)^{5/2}.
\end{equation}
This expression is consistent with the results of \citet{2023ApJ...947...32C} (who use a modified Lane--Emden equation; \citealt{1935MNRAS..95..226C}) and with \citet{2010MNRAS.409L..25Z}. We take the onset of mass loss to occur for $\beta>\beta_{0}$ with $\beta_{0}=0.5$, for which the Roche-lobe radius may be written as
\begin{equation}
    R_{\mathrm{roche}} = \beta_{0} R_{\mathrm{p}}\left(\frac{M_{\mathrm{h}}}{M_{\star}}\right)^{-1/3} = \left(\frac{\beta_{0}}{\beta}\right)R_{\star}.
    \label{eqt:roche lobe}
\end{equation}
Here $R_{\mathrm{p}}$ is the pericentre radius and $\beta=R_{\mathrm{T}}/R_{\mathrm{p}}$. Substituting Equation~\ref{eqt:roche lobe} into $\Delta = R - R_{\mathrm{roche}}$, we obtain
\begin{equation}
    \frac{\Delta}{R} = 1 - \left(\frac{R_{\mathrm{p}}}{2R_{\mathrm{T}}}\right)\left(\frac{M_{\star}}{M_{0}}\right)^{2/3}
    \sqrt{\frac{1 - (M_{0}/M_{\mathrm{Ch}})^{4/3}}{1 - (M_{\star}/M_{\mathrm{Ch}})^{4/3}}}.
    \label{eqt:roche_lobe_overflow}
\end{equation}

\section{Simulation results} \label{sec:Simulation results}
In this section we present two aspects of our simulation results: (i) the mass fallback rate following a single encounter, and (ii) the element-specific fallback curves that encode the WD's layered composition. 

\subsection{Mass Fallback Rate} \label{sec:Mass fallback Rate}
\begin{figure*}[htbp]
\centering
\gridline{\fig{ 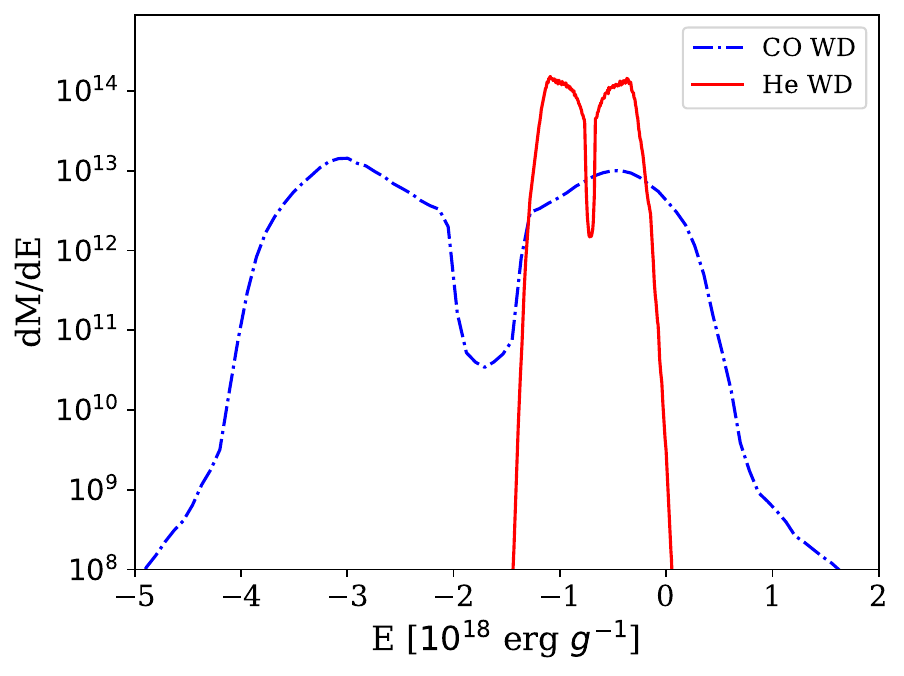}{0.48\textwidth}{}
          \fig{ 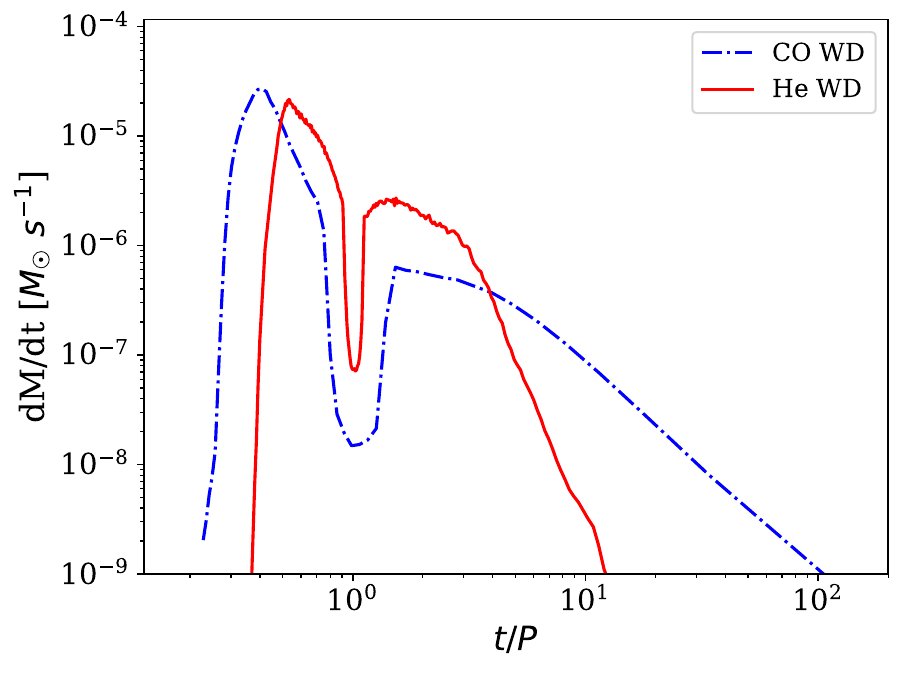}{0.48\textwidth}{}
         }
\caption{Specific orbital-energy distribution (left) and fallback rate (right) for a $0.15\,M_{\odot}$ He WD and a $0.6\,M_{\odot}$ CO WD after an encounter with a $10^{4}\,M_{\odot}$ BH with $e=0.9$ and $\beta=0.7$. In the right panel, the time axis is normalized by the orbital period, $P=4838\,\mathrm{s}$ (He) and $P=1336\,\mathrm{s}$ (CO). \label{fig:fallback}}
\end{figure*}

Figure~\ref{fig:fallback} shows the specific orbital-energy distribution and the corresponding fallback rate for a $0.15\,M_{\odot}$ He WD and a $0.6\,M_{\odot}$ CO WD after an encounter with a $10^{4}\,M_{\odot}$ BH with $e=0.9$ and $\beta=0.7$. Both $dM/dE$ and $\dot{M}(t)$ exhibit a double-peaked structure: the first peak corresponds to the leading (``front'') arm of the debris stream and the second peak to the trailing (``back'') arm. Because the leading arm is more tightly bound (more negative specific energy), it returns to pericentre first, followed by the remnant core and then the trailing arm.

The CO WD exhibits a broader energy spread, with $\Delta\epsilon$ about $3.5$ times larger than that of the He WD. For fixed $\beta\equiv R_{\mathrm{T}}/R_{\mathrm{p}}$, the more compact CO WD has a smaller physical pericentre distance $R_{\mathrm{p}}$, implying a larger characteristic energy scale ($\Delta\epsilon\propto R_{\mathrm{p}}^{-2}$) and a stronger tidal perturbation that more efficiently redistributes orbital energy, widening $dM/dE$. By contrast, the less compact He WD has a larger $R_{\mathrm{p}}$ and a correspondingly narrower energy distribution.

Using Equation~\ref{eqt:fallback_timescale}, we estimate fallback timescales of $t_{\mathrm{fb}}\approx 1474\,\mathrm{s}$ (He WD) and $t_{\mathrm{fb}}\approx 269\,\mathrm{s}$ (CO WD), implying minute-to-hour timescale flares. The earlier fallback of the CO WD reflects its more tightly bound debris. Although the precise onset of fallback depends on hydrodynamic details, the analytic estimates are consistent with the simulations and support the choice $\xi=3.5$ from \citet{2023ApJ...947...32C}.

The diagnostics above quantify when debris returns to pericentre, but they do not reveal its composition. Because WD envelopes and cores occupy distinct radial shells with different characteristic binding energies, tidal stripping should imprint a time-dependent compositional signature on both $dM/dE$ and $\dot{M}(t)$. We therefore use the tracer abundances in our simulations to compute element-specific energy distributions and fallback rates.

\begin{figure*}[htbp]
\centering
\gridline{\fig{ 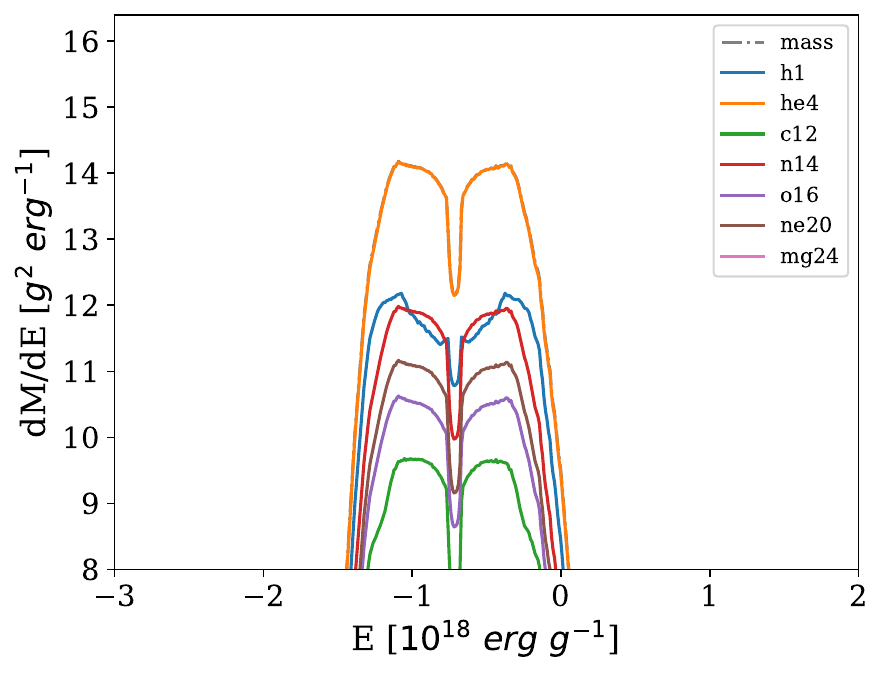}{0.48\textwidth}{}
          \fig{ 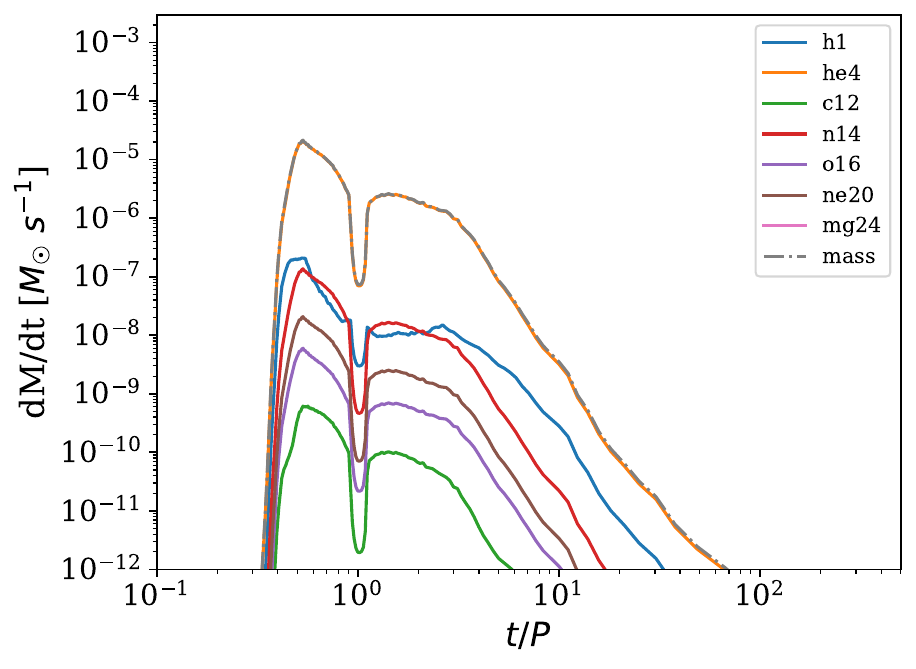}{0.48\textwidth}{}
         }
\caption{Element-specific energy distribution (left) and fallback rate (right) for a $0.15\,M_{\odot}$ He WD after an encounter with $\beta=0.7$ (same simulation as Figure~\ref{fig:fallback}). \label{fig:element_specific_He_fallback}}
\end{figure*}

\subsection{Compositional Mass Fallback Rate} \label{sec:Composition dependence of Mass fallback Rate}
\begin{figure*}[htbp]
\centering
\gridline{\fig{ 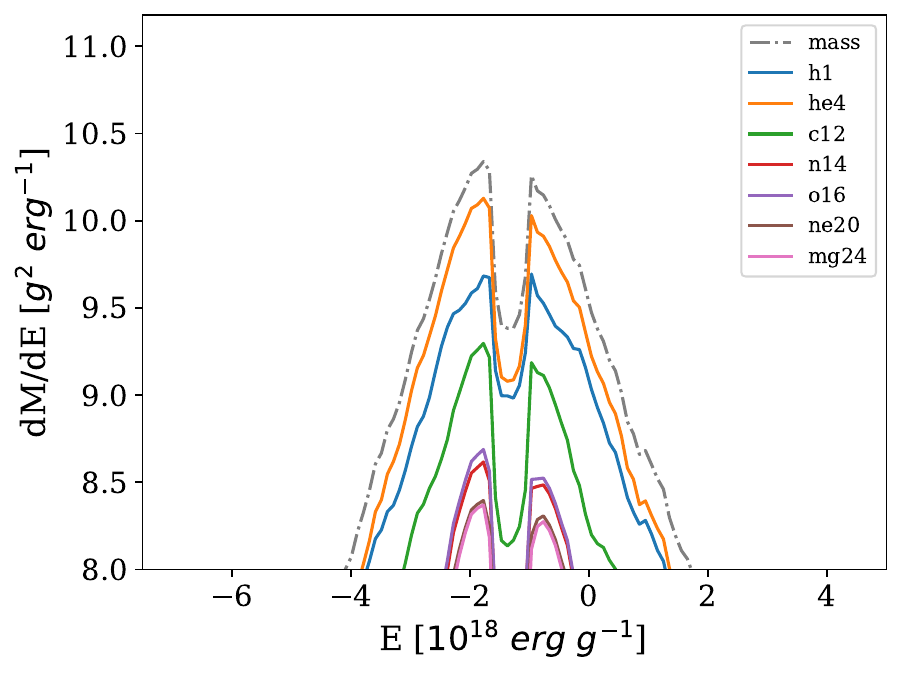}{0.48\textwidth}{}
          \fig{ 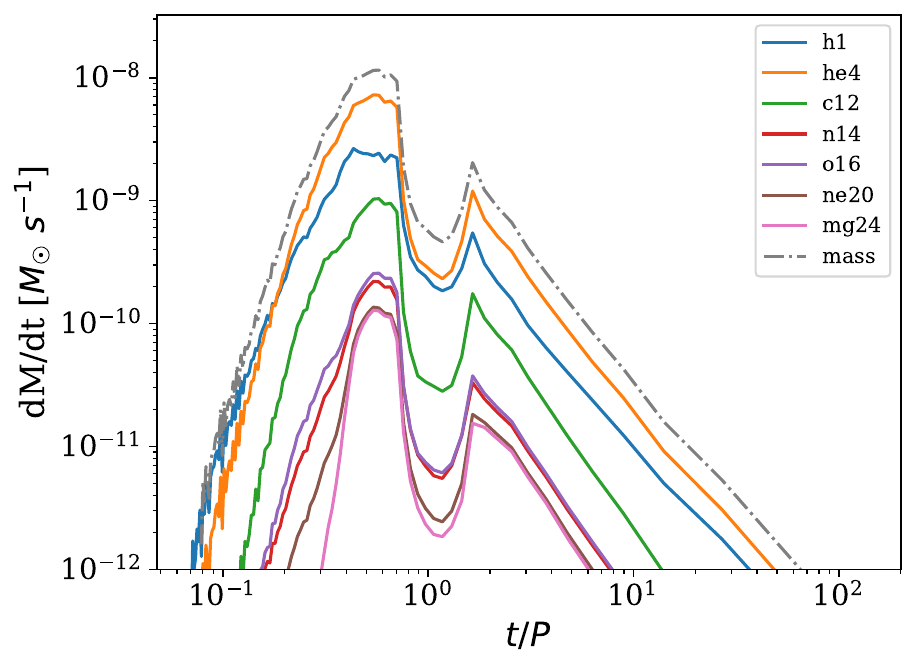}{0.48\textwidth}{}
         }
\gridline{\fig{ 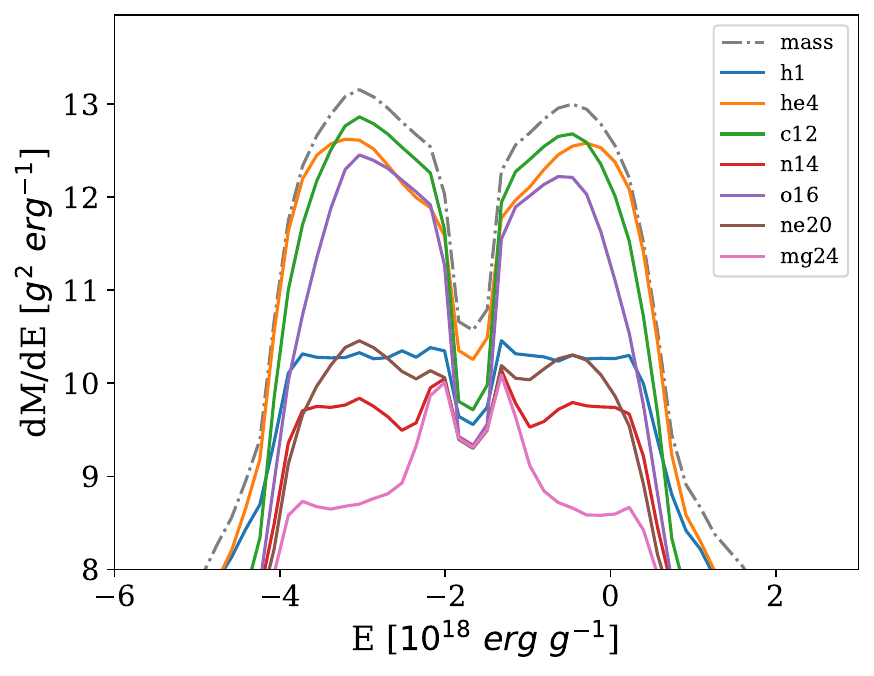}{0.48\textwidth}{}
          \fig{ 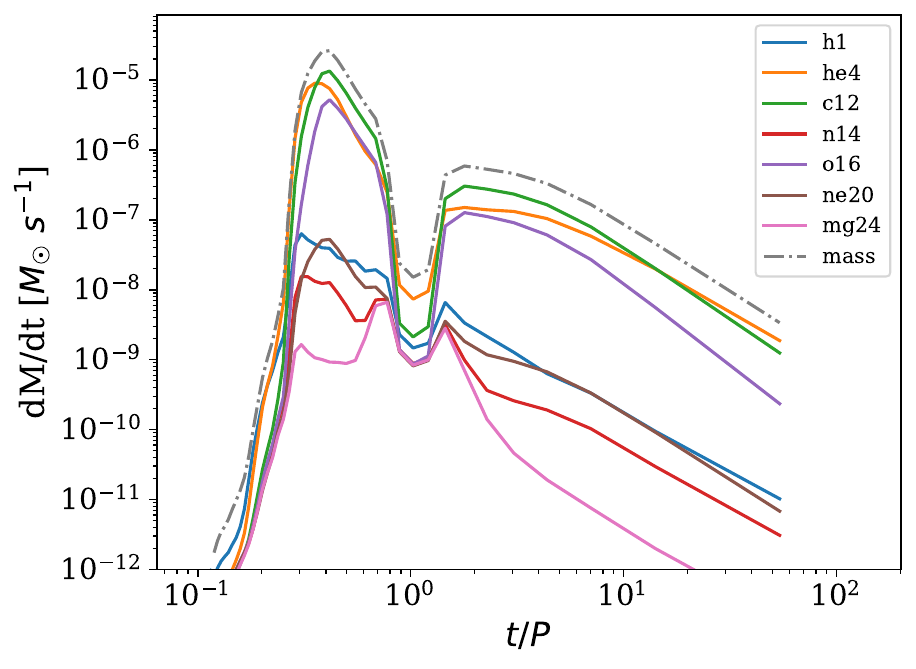}{0.48\textwidth}{}
         }
\gridline{\fig{ 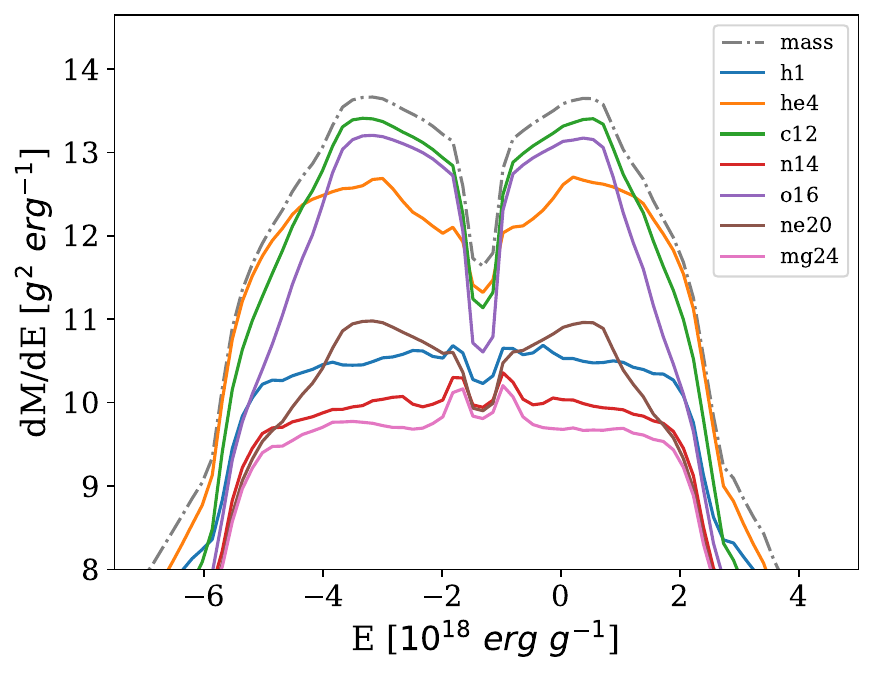}{0.48\textwidth}{}
          \fig{ 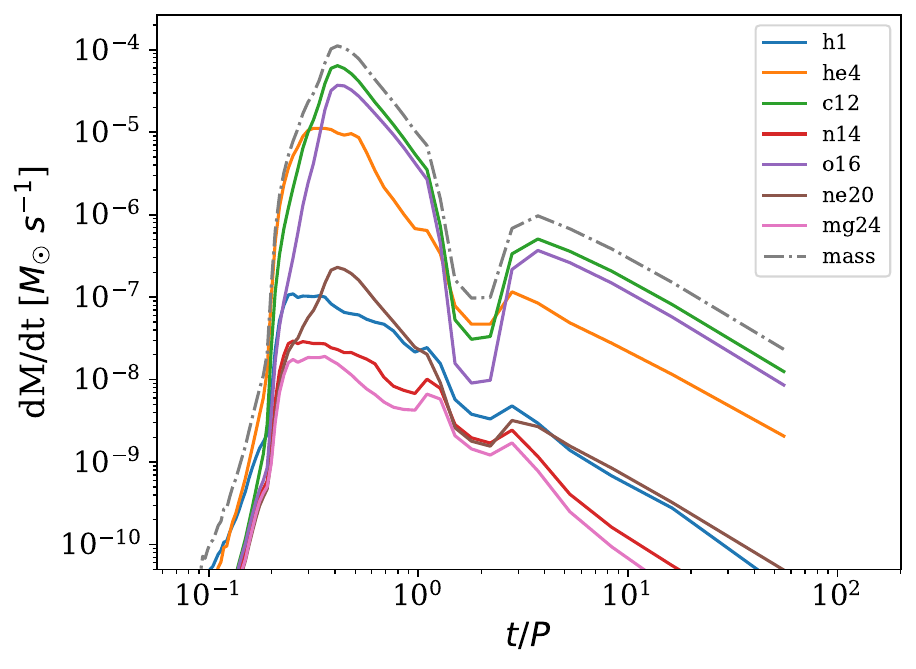}{0.48\textwidth}{}
         }
\caption{The element-specific binding energy distribution (left panels) and fallback rate (right panels) of a $0.6 M_{\odot}$ CO WD after the tidal encounter with $\beta = 0.55, 0.7, 0.8$, from top to bottom panels. \label{fig:element_specific_fallback}}
\end{figure*}

Figure~\ref{fig:element_specific_He_fallback} shows $\mathrm{d}M/\mathrm{d}E$ for the $0.15\ M_\odot$ He WD disrupted at $\beta = 0.7$, while Figure~\ref{fig:element_specific_fallback} shows the same for a $0.6\ M_\odot$ CO WD at $\beta = 0.55, 0.7, 0.8$. In all cases, each element follows the same double-peaked and symmetric shape as the overall mass fallback rate. Here, we examine how the relative contribution of each element varies across this structure, reflecting the different spatial distributions of core and envelope material.

\citet{2017ApJ...841..132L} showed that, for the full disruption of a low-mass He WD with a H envelope on a parabolic orbit, the fallback composition depends on $\beta$. In a grazing encounter, only the H-rich envelope is stripped and dominates the fallback. In contrast, a deeper encounter that penetrates the core produces a compositional transition from envelope to core (H-to-He) before $t_{\text{peak}}$.

In Figure~\ref{fig:element_specific_He_fallback}, our simulation of the He WD with $\beta = 0.7$ instead shows a fallback dominated by He4 throughout, with no H-to-He transition before or after $t_{\text{peak}}$. Following the $\beta$-dependence reported in \citet{2017ApJ...841..132L}, this indicates that the encounter is sufficiently deep for both H1 and He4 to fall back simultaneously, producing a core-dominated fallback without a resolvable transition, or one that occurs too early to be distinguished from the initial fallback noise. This result extends Figure 10 of \citet{2017ApJ...841..132L} to a deeper encounter, in the sense of penetration relative to the structure of our He WD, although our actual value of $\beta$ is smaller than that used in their right panel.

In Figure~\ref{fig:element_specific_fallback}, we extend this analysis to a $0.6\ M_\odot$ CO WD with He and H envelopes, constructed in \texttt{MESA}, under partial disruption with $\beta = 0.55, 0.7, 0.8$. The fallback is primarily composed of H1, He4, C12, and O16, with the abundance of heavier elements increasing with $\beta$. For $\beta = 0.55$, the envelope element He4 dominates the fallback peak, while for $\beta = 0.7$ and $0.8$, the core element C12 dominates, with O16 becoming increasingly significant and expected to dominate for even larger $\beta$.
The composition also evolves throughout a single fallback event:
\begin{itemize}
\item For $\beta = 0.55$ (envelope-only stripping), H1 dominates at very early times and transitions to He4 at $\sim 0.3 t_{\text{peak}}$, with little C12 or O16 present.
\item For $\beta = 0.7$ (shallow core penetration), the He4-to-C12 transition occurs at $\sim 0.9 t_{\text{peak}}$.
\item For $\beta = 0.8$ (deeper core penetration), the He4-to-C12 transition is stronger and occurs earlier, at $\sim 0.75 t_{\text{peak}}$.
\end{itemize}

This establishes two trends, the He-to-C transition occurs progressively earlier, and its strength increases, with larger $\beta$. To quantify transition strength, we define the ratio of the peak fallback rate between two elements,
\begin{equation}
     q_{\text{h, he}} = \frac{(dM/dt)_{\text{he, peak}}}{(dM/dt)_{\text{h, peak}}},
\end{equation}

\begin{equation}
     q_{\text{he, c}} = \frac{(dM/dt)_{\text{c, peak}}}{(dM/dt)_{\text{he, peak}}}. 
\end{equation}

We find $q_{\text{h, he}} \simeq 2.74$ for $\beta = 0.55$, and $q_{\text{he, c}} \simeq 1.51$ and $5.76$ for $\beta = 0.7$ and $0.8$, respectively. Since these transitions trace the layered composition of the WD, we expect an analogous C12-to-O16 transition to emerge for $\beta > 0.8$. 

In Figure~\ref{fig:abundance_evolution}, we show the complete evolution of different elements up to $60 t_{\rm peak}$. A log-log version of this figure is presented in \ref{Appendix B}. Additionally, we plot the abundance evolution of the CO WD relative to solar abundances in \ref{Appendix A}. 

These compositional changes should directly imprint on the observed light curves and spectral line ratios \citep{2016ApJ...827....3R,2017ApJ...841..132L,2018ApJ...857..109G,2019ApJ...882L..25L}. \citet{2017ApJ...841..132L} predicted, without explicit spectral calculations, a smooth H-to-He transition in line diagnostics with increasing $\beta$ for He WD disruptions, with the grazing-encounter case being purely H-powered. For CO WDs, \citet{2011ApJ...726...34C} calculated emission lines powered by photoionization from unbound debris for $\beta = 1$ and $\beta = 3$, finding $\mathrm{CIV} \ \lambda1549$ to be strongest at early times and $\mathrm{[OIII]} \ \lambda5007$ at later times. However, the existing models assume a fixed composition throughout the CO WD, with elements uniformly distributed, and focus on deep encounters, neglecting the stratified structure of realistic WDs.  

We extend this picture to partial disruptions of stratified CO WDs. By analogy with \citet{2017ApJ...841..132L}, we predict a smooth He-to-C transition in line diagnostics with increasing $\beta$. A grazing encounter will yield a helium-dominated spectrum (the H lines in early times will be suppressed by He lines, following \citealt{2016ApJ...827....3R}), while increasing $\beta$ raises the C/He line ratio in earlier times, tracking $q_{\text{he, c}}$, with O present throughout. In sufficiently strong encounters penetrating deeply in the core, we further suggest a C-to-O transition, with the O/C line ratio $q_{\text{c, o}}$ increasing with $\beta$.

Based on this analogy, we predict that CO WD disruption spectra would show strong He and C, absent N, and moderate O, with He line strength decreasing and O line strength increasing with $\beta$. The absence of N is a key discriminant. As the disruptions of evolved MS star progenitors instead show strong N, weak C, and moderate O \citep{2019ApJ...882L..25L}. This compositional signature thus offers a way to distinguish the disruption of compact objects from that of evolved stars in future UV spectroscopic TDE samples. Though we emphasize this is a compositional expectation, not a radiative transfer calculation, and the resulting line strengths would depend on ionization and optical depth effects we have not modeled here. 

\textit{Caveat:} the above assumes a short viscous timescale, and a rapid circularization, where the light curve tracks the mass fallback rate directly, and the abundance features in the spectrum tracks $\dot{M}_X(t) = X \dot{M}(t)$ almost instantaneously, where $X$ is the abundance of the element. If flares instead undergo a lot of reprocessing by the accretion disk wind, the relevant composition is that of the accumulated disk material ${M}_X$, integrated over all prior fallback as shown in Figure~\ref{fig:cumulative_abundance_evolution}, rather than the instantaneous fallback composition. From \citet{2019ApJ...872..151M}, most of the observed TDE light curves track the fallback rate without significant viscous delays, implying a short viscous timescale, therefore our assumption is valid. 

\section{Theoretical Predictions for Repeating TDEs}\label{sec:theory_repeating_tdes}
Although we do not simulate multiple encounters directly, we can use our single-passage results to make predictions for repeating TDEs. Over successive pericentre passages, the WD is stripped from the outside in, progressively exposing deeper compositional layers. In the previous section, we used numerical simulations to relate the stripped-layer depth (and its composition) to $\beta$; here we extrapolate those results to predict how the fallback composition evolves from passage to passage.

This evolution also depends on how many passages the WD remnant survives. Below, we estimate the mass stripped per encounter and the maximum number of survivable passages for a WD of given mass and composition using the analytic framework developed in this section.

\subsection{Orbital Evolution}\label{sec:Orbit}
As the WD orbits the BH, it loses angular momentum to GW radiation \citep{1964PhDT........51P}. The semi-major axis $a$ and eccentricity $e$ therefore evolve slowly, with orbit-averaged rates
\begin{equation}
    \dot{a} \simeq -\frac{64}{5} \frac{G^{3} M_{\star} M_{\mathrm{h}}^{2}}{c^{5} a^{3} (1-e^{2})^{7/2}} \left( 1 + \frac{73}{24} e^{2} + \frac{37}{96} e^{4} \right),
\end{equation}
\begin{equation}
    \dot{e} \simeq -\frac{304}{15} e \frac{G^{3} M_{\star} M_{\mathrm{h}}^{2}}{c^{5} a^{4} (1-e^{2})^{5/2}} \left( 1 + \frac{121}{304} e^{2} \right),
\label{eqt:e_evolve_GW}
\end{equation}
so that after one orbital period $P$ we estimate $a_{\mathrm{new}} \simeq a + P\dot{a}$ and $e_{\mathrm{new}} \simeq e + P\dot{e}$. 
The pericentre distance on the next orbit is then $R_{\mathrm{p,new}} = a_{\mathrm{new}}(1-e_{\mathrm{new}})$. 

To account for orbital changes due to tidal excitation and mass transfer, \citet{2024ApJ...977...80C} found that the change in a WD's orbital energy following a partial tidal disruption can be either positive, due to mass loss, or negative, due to tidal oscillations, depending on $\beta$. The relation between $\Delta \epsilon_{\rm orb}$ and $\beta$ is given by Equation $20$ in \citet{2024ApJ...977...80C}:
\begin{equation}
    \Delta \epsilon_{\rm orb, 0} \sim \frac{GM_{\star}}{R_{\star}}(\frac{\Delta M}{M_{\rm rem}} - \beta^6T_2), 
\label{eqt:orbital_energy_change}    
\end{equation} 
where $M_{\rm rem}$ is the remnant mass, and $T_2$ depends on the stellar structure and oscillatory mode. Since the calculations in \citet{2024ApJ...977...80C} assume a $\gamma = 5/3$ polytrope with $M_{\star}=M_{\odot}$ and $R_{\star}=R_{\odot}$, and $\Delta \epsilon_{\rm orb} \propto GM_{\star}/R_{\star}$, we can scale Equation~\ref{eqt:orbital_energy_change} using a dimensionless parameter $\eta = (M_{\star}/ R_{\star}) / (M_{\odot} / R_{\odot})$. The orbital energy change of any WD is then given by $\Delta \epsilon_{orb} = \eta \Delta \epsilon_{orb, 0}$, and the corresponding change in semi-major axis is $\Delta a = 2a^2\Delta \epsilon_{\rm orb} / GM_{\rm h}$. 

Adopting their results and combining with the effect of GW emission, we can calculate the orbital change of a WD per orbit and the long-term evolution of a WD-IMBH system undergoing repeated tidal stripping.

\subsection{Repeating Mass Loss Process}\label{sec:Repeating_Mass_Loss}
Starting from the initial WD mass and radius given by \texttt{MESA}, we iteratively compute the stripped mass $\delta M$ at each pericentre passage while updating $M_{\star}$, $a$, $e$, and $R_{\mathrm{p}}$. We stop the iteration when the WD is fully disrupted, defined by $\delta M/M_{\star} \geq 1$ or $R_{\mathrm{p}} < R_{\mathrm{T}}$. This semi-analytic model lets us follow the long-term evolution over many passages that a single hydrodynamic simulation cannot practically capture.

\begin{figure*}[htbp]
\centering
\gridline{\fig{ 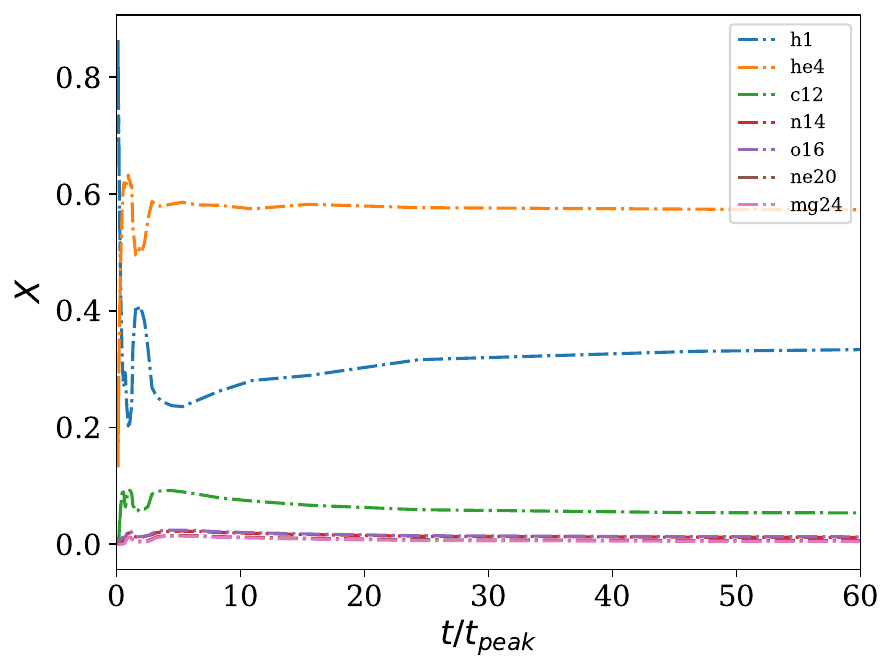}{0.48\textwidth}{}
          \fig{ 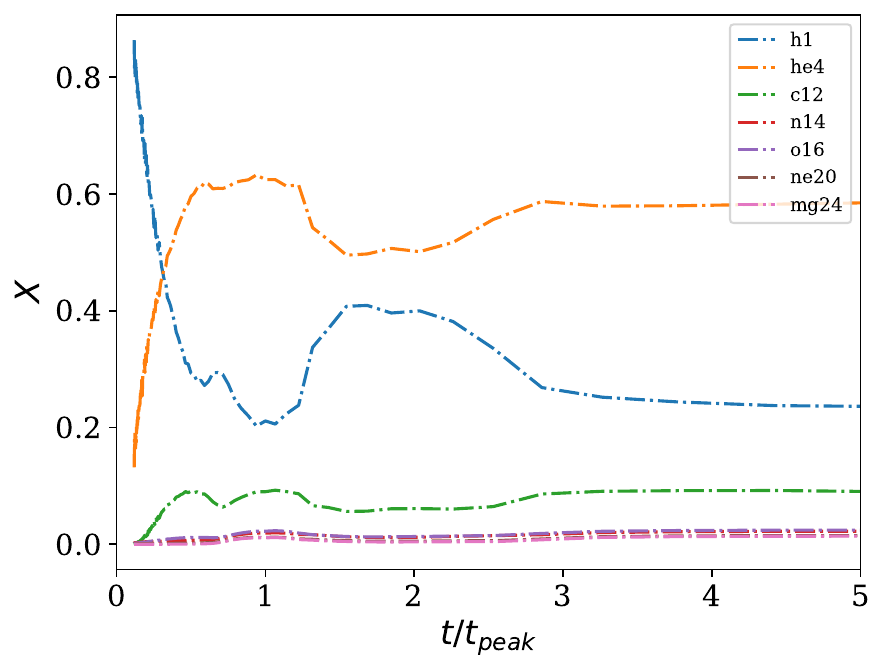}{0.48\textwidth}{}
         }
\gridline{\fig{ 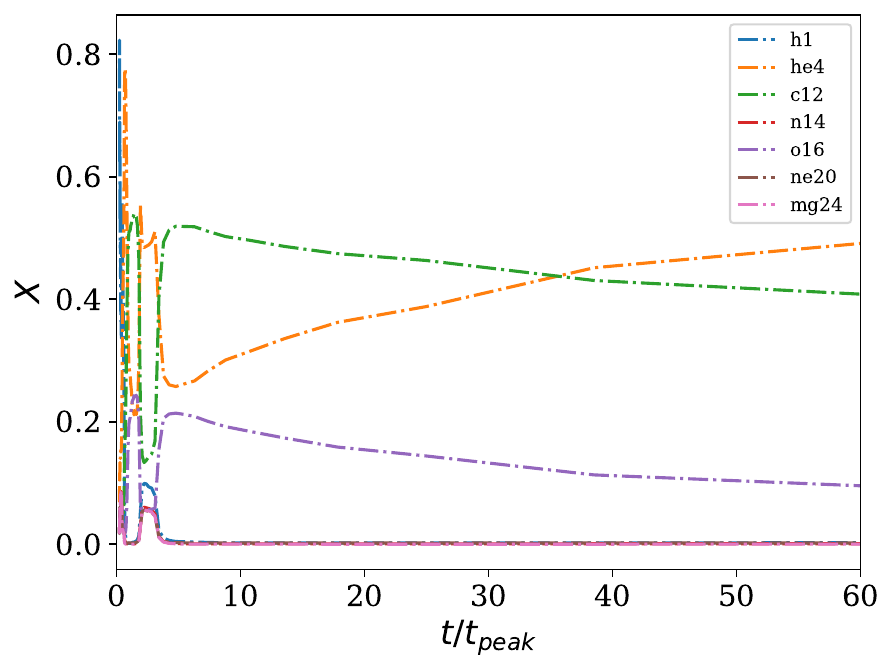}{0.48\textwidth}{}
          \fig{ 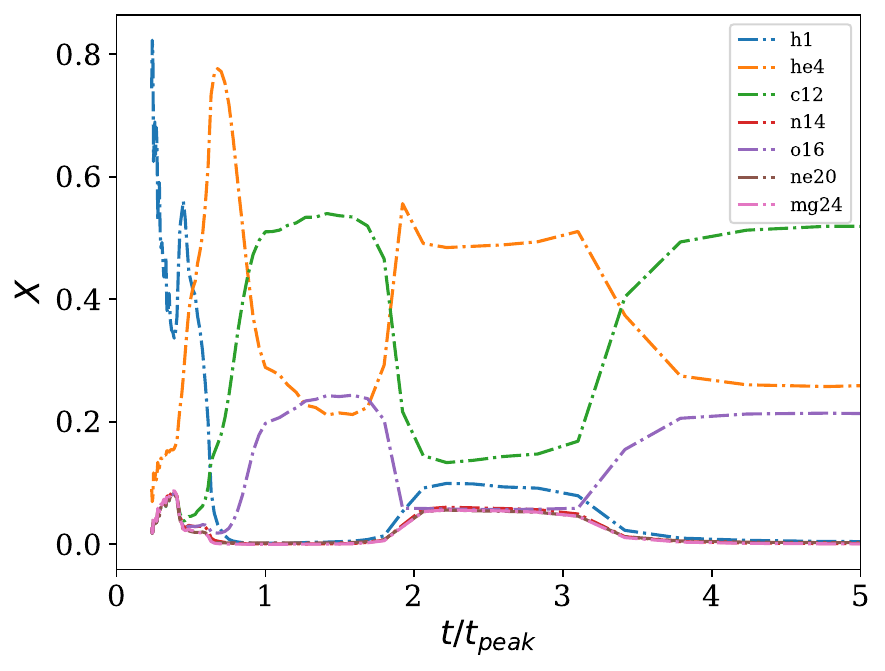}{0.48\textwidth}{}
         }
\gridline{\fig{ 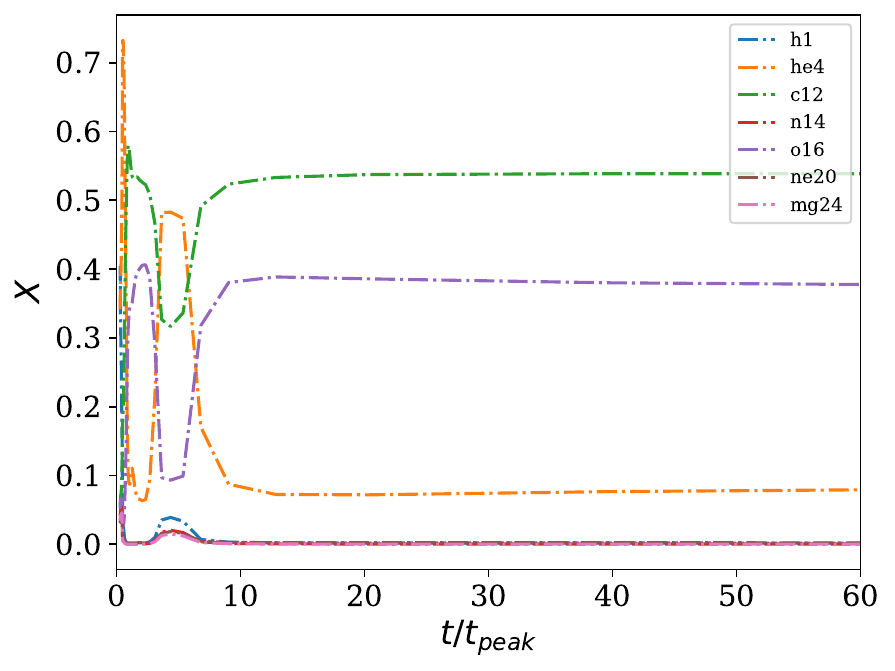}{0.48\textwidth}{}
          \fig{ 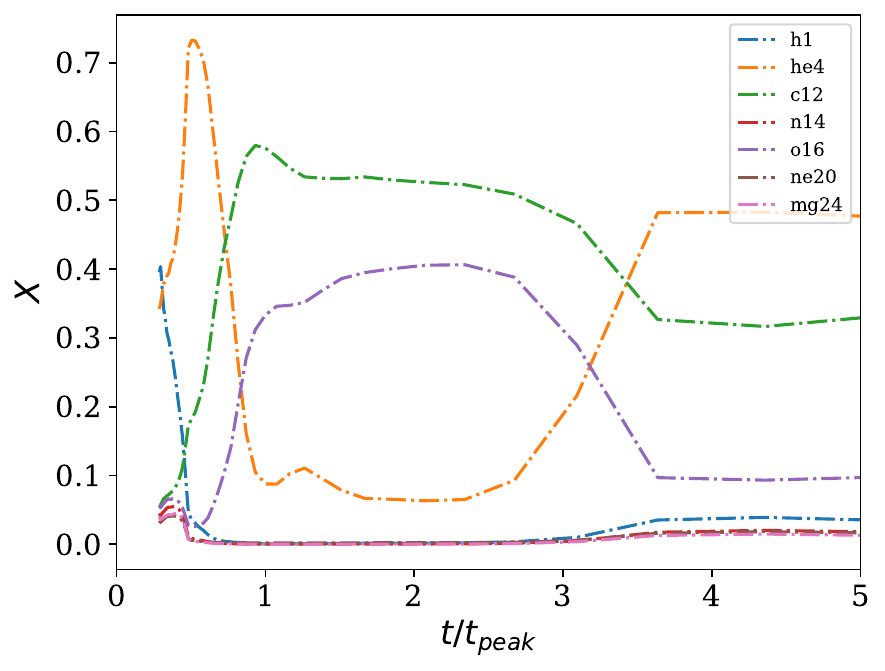}{0.48\textwidth}{}
         }
\caption{Left panels: abundance of different elements as a function of time for disruption of a $0.6\,M_{\odot}$ CO WD in the same encounters as Figure~\ref{fig:element_specific_fallback} (top to bottom: $\beta = 0.55, 0.7, 0.8$). Right panels: zoom-in at early times ($t < 2t_{\mathrm{peak}}$).\label{fig:abundance_evolution}}
\end{figure*}

\begin{figure*}[htbp]
\centering
\gridline{\fig{ 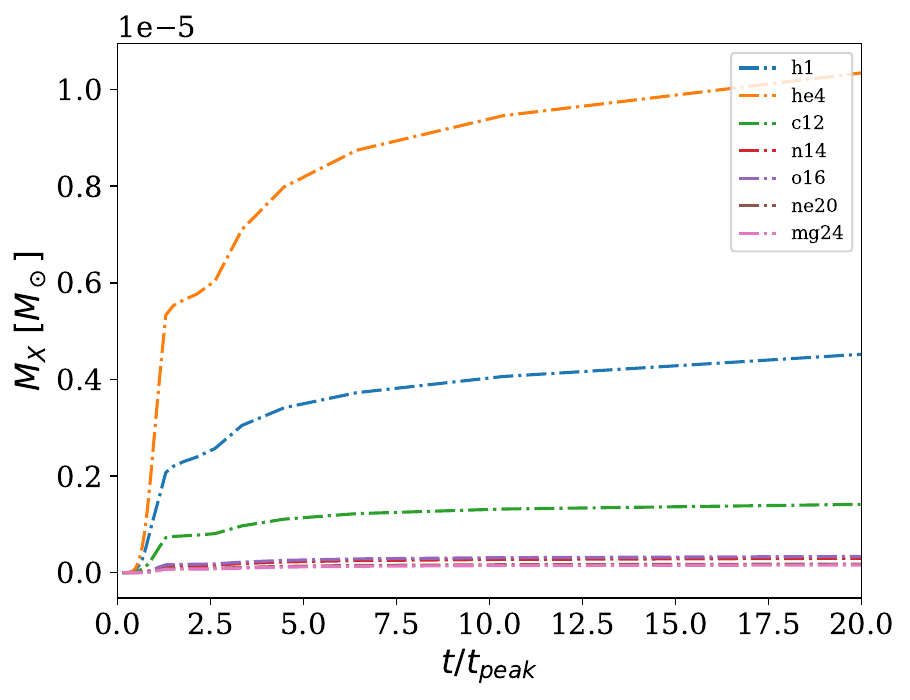}{0.48\textwidth}{}
          \fig{ 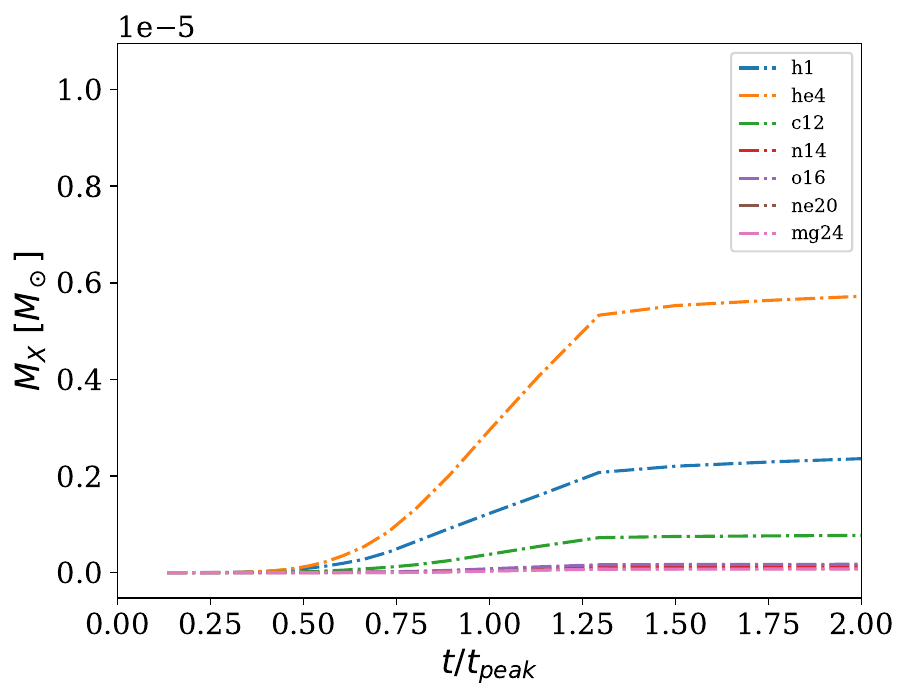}{0.48\textwidth}{}
         }
\gridline{\fig{ 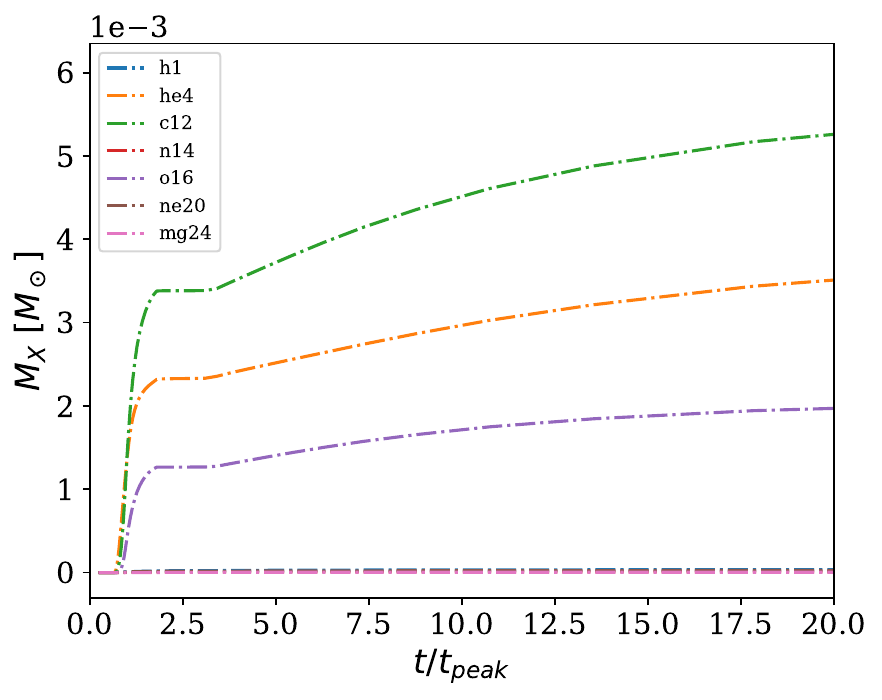}{0.48\textwidth}{}
          \fig{ 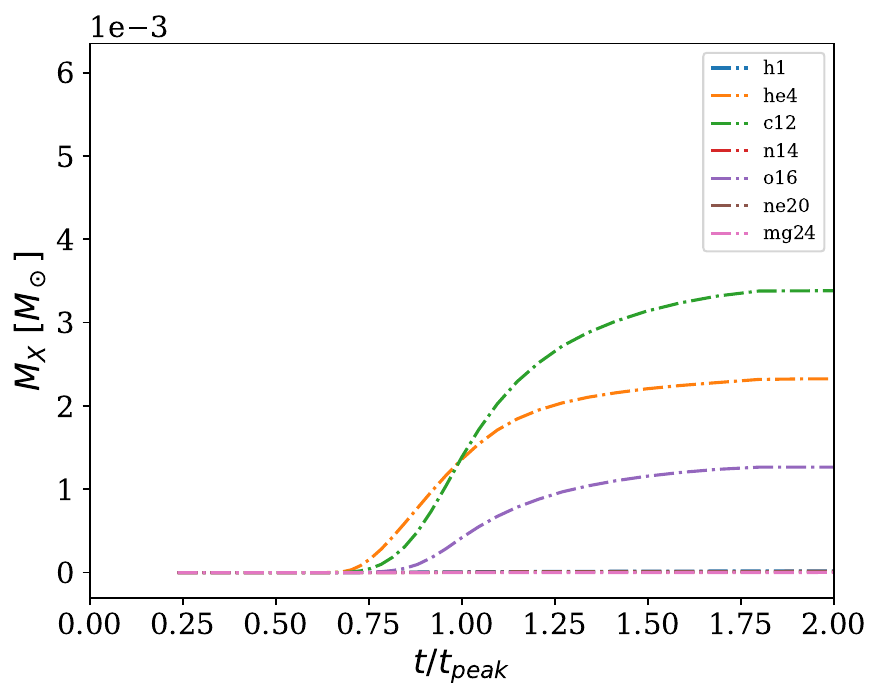}{0.48\textwidth}{}
         }
\gridline{\fig{ 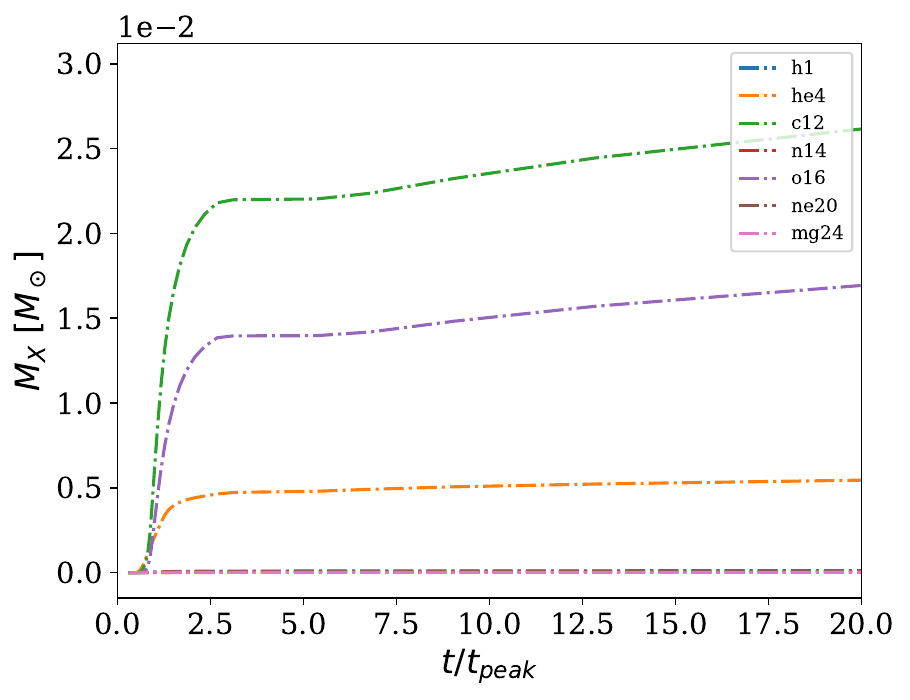}{0.48\textwidth}{}
          \fig{ 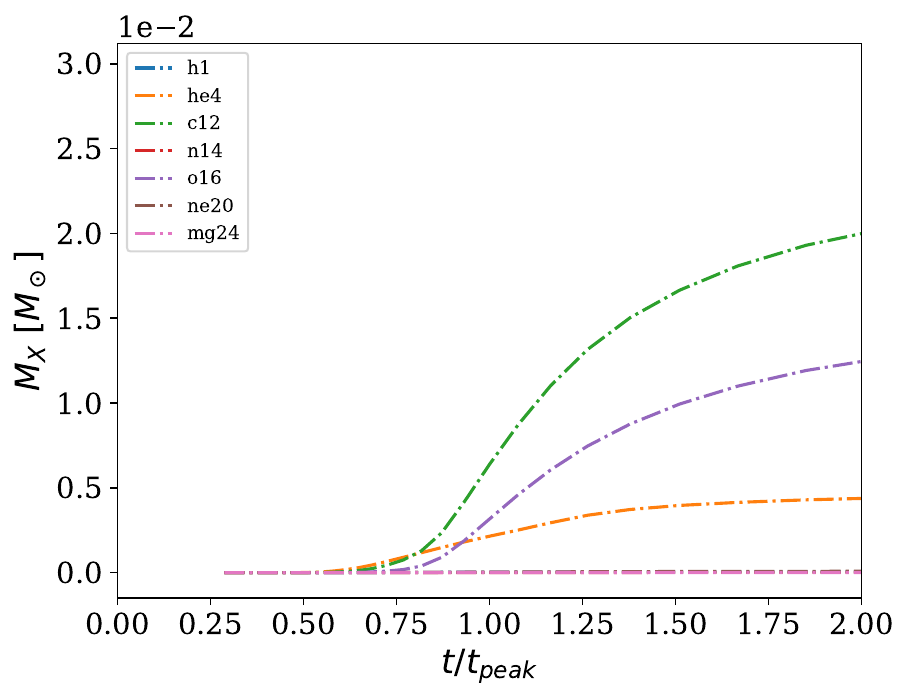}{0.48\textwidth}{}
         }
\caption{Left panels: cumulative fallback mass in different elements as a function of time for disruption of a $0.6\,M_{\odot}$ CO WD in the same encounters as Figure~\ref{fig:element_specific_fallback} (top to bottom: $\beta = 0.55, 0.7, 0.8$). Right panels: zoom-in at early times ($t < 2t_{\mathrm{peak}}$).\label{fig:cumulative_abundance_evolution}}
\end{figure*}

During tidal stripping, when the tidal force on the WD surface overcomes self-gravity, the star overflows its Roche lobe and begins to lose mass \citep{2010MNRAS.409L..25Z}. This process can repeat at each pericentre passage until the star is fully disrupted or the remnant is ejected from the WD--IMBH system \citep{2024ApJ...977...80C}.

\begin{figure*}[htbp!]
\centering
\gridline{\fig{ 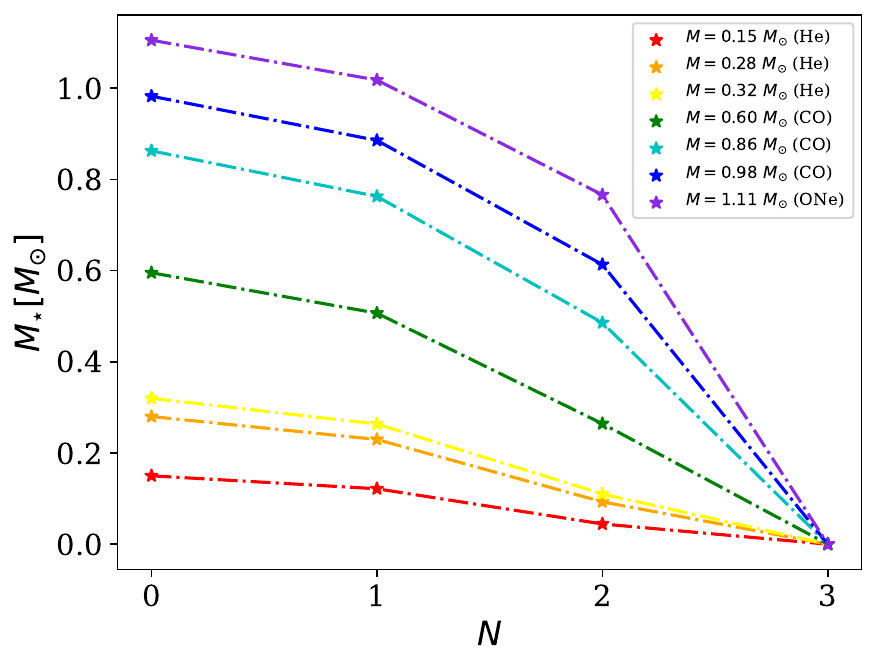}{0.48\textwidth}{}
          \fig{ 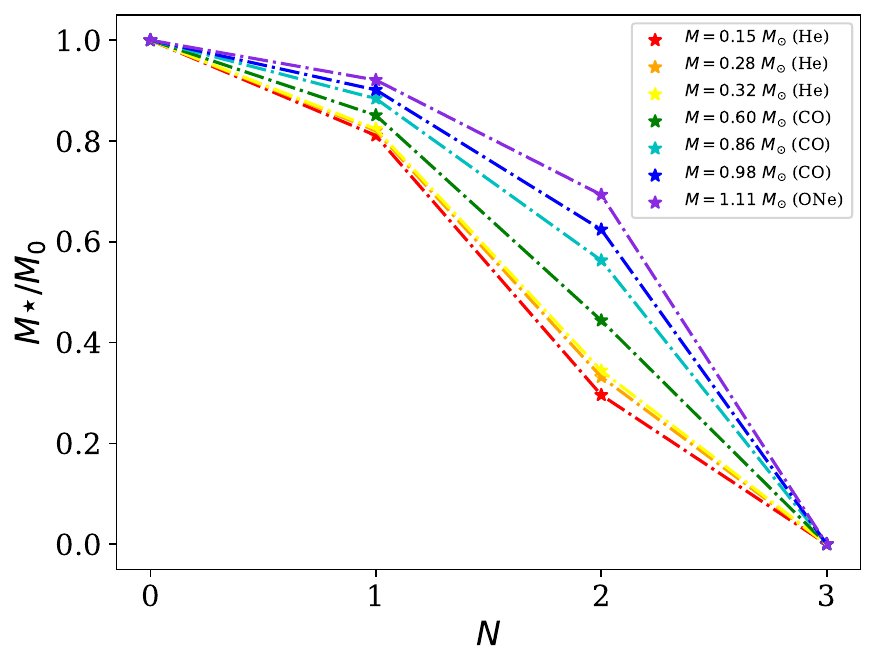}{0.48\textwidth}{}
         }
\caption{WD remnant mass versus passage number for seven WDs constructed with \texttt{MESA}, with initial orbital parameters $e_0 = 0.9$ and $\beta_0 = 0.7$. Left: remnant mass. Right: remnant mass normalized by the initial WD mass.\label{fig:number_of_passages}}
\end{figure*}

Figure~\ref{fig:number_of_passages} shows the WD remnant mass as a function of passage number for seven WDs with different initial masses and compositions (with $M_0$ and $R_0$ from \texttt{MESA}). We adopt initial orbital parameters $e_0 = 0.9$ and $\beta_0 = 0.7$ and iterate until full disruption. The remnant mass of He WDs drops steeply over successive orbits, while the fractional mass loss per passage, $\delta M/M_{\star}$, is smaller for denser WDs (CO and ONe) due to their higher compactness. For $\beta_0 = 0.7$, all WDs are completely disrupted by the third passage. For deeper encounters ($\beta_0 \gtrsim 1$), disruption can occur in a single passage.

We also compare the analytically predicted stripped mass in a single encounter for a $0.15\,M_{\odot}$ He WD and a $0.6\,M_{\odot}$ CO WD with our simulation results. The agreement is good, indicating that the framework in Equations~(\ref{eqt:mass_radius_relationship})--(\ref{eqt:roche_lobe_overflow}) provides a reasonable approximation for mass loss during tidal stripping.

\begin{figure*}[ht!]
\centering
\plotone{ 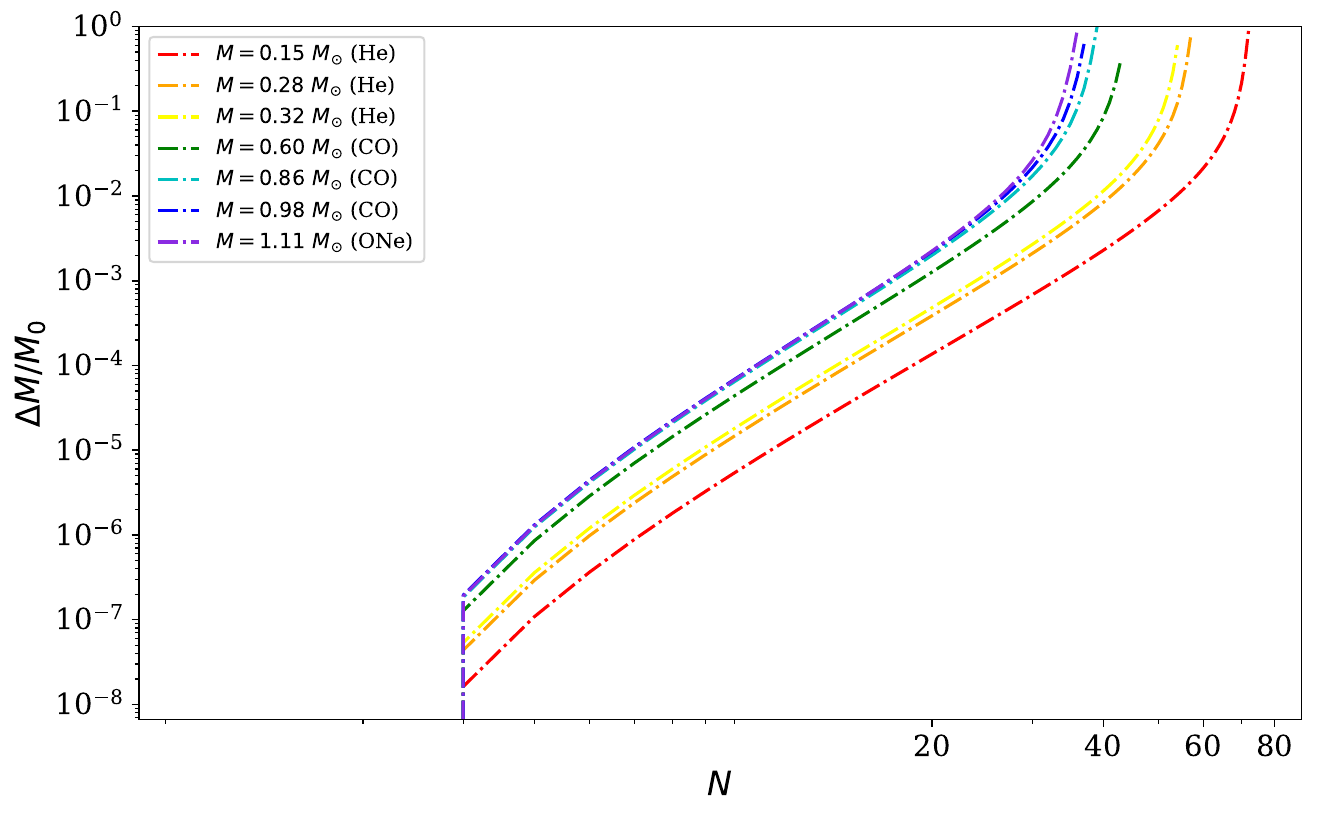}
\caption{Ratio of cumulative mass loss to the initial mass of seven WDs versus passage number for repeating tidal stripping on an orbit with $e_0 = 0.9$ and $\beta_0 = 0.5$. From this figure we infer maximum passage numbers spanning $\sim 30 - 80$, depending on the initial WD mass.\label{fig:max_number_of_passages}}
\end{figure*}

To estimate the maximum survivable number of passages, we consider an initial impact parameter $\beta_0 = 0.5$, close to the onset of mass loss (smaller $\beta_0$ generally implies more survivable passages). For $\beta < 0.4$, the tidal force is too weak to produce a TDE \citep{2024ApJ...977...80C}. Even for grazing encounters with $0.4 \leq \beta \leq 0.5$, the accreted mass is $\lesssim 10^{-6}\,M_{\odot}$ \citep{2024ApJ...977...80C}, which may be insufficient to power detectable electromagnetic emission. Figure~\ref{fig:max_number_of_passages} shows the resulting $N_{\max}$ for the same seven WDs with $e_0 = 0.9$ and $\beta_0 = 0.5$. 

The cumulative fractional mass loss as a function of passage number can be divided into three phases. In the first phase, stripping is negligible over the first few orbits. In the second phase, the orbit shrinks slowly and the WD undergoes repeated, brief Roche-lobe overflow episodes, producing a gradual increase in cumulative mass loss that approximately follows a power law,
\begin{equation}
    \frac{M_0 - M_{\star}}{M_0} \propto N^{\zeta}.
\end{equation}
We find $\zeta \simeq 4.32-4.81$, where $\zeta$ is larger for denser WDs, implying accelerating mass loss: later passages remove substantially more mass than early passages as the Roche-lobe overflow depth $\Delta$ grows from orbit to orbit \citep{2010MNRAS.409L..25Z} (through the dependence on decreasing $R_{\mathrm{p}}$ and $M_{\star}$ in Equation~\ref{eqt:roche_lobe_overflow}). In the third phase, the WD reaches a critical point and mass loss increases rapidly, leading to full disruption within a few orbits.

All seven WDs reach this critical point when the cumulative fractional mass loss is $\sim 5\times 10^{-2}$ in our $\beta_0 = 0.5$ calculations. This suggests that once a small critical fraction of the mass has been removed, the remnant becomes increasingly vulnerable due to structural changes induced by previous encounters. For comparison, \citet{2010MNRAS.409L..25Z} found a similar transition with $(\Delta M/M_0)_{\mathrm{crit}} \sim 10^{-2}$ for different eccentricities at $\beta_0 = 0.5$ in an encounter with a $10^{5}M_{\odot}$ BH. Extending our calculations over a range of $\beta_0$ and $e_0$, we find that $(\Delta M/M_0)_{\mathrm{crit}}$ is approximately independent of eccentricity. However, the well-defined $(\Delta M/M_0)_{\mathrm{crit}}$ appears to be a special case at $\beta_0 = 0.5$. With increasing $\beta_0$, the cumulative mass loss curve becomes flatten, and the WD undergoes full disruption within a small number of passages, without exhibiting a transition to a runaway mass loss. 

The maximum number of passages varies among WDs with different internal structures. For the weaker encounter considered here ($\beta_0 = 0.5$), less dense WDs can survive more passages than denser ones. For a fixed $\beta$, denser (and smaller) WDs have smaller pericentre distances because $R_{\mathrm{p}} \propto R_{\mathrm{T}} \propto M_{\star}^{-1/3}R_{\star}$, and WDs follow an inverse mass--radius relation. The shorter pericentre distances lead to larger GW-driven orbital changes per passage, causing the orbits of CO and ONe WDs to shrink more rapidly than those of He WDs. Additionally, denser WDs experience larger changes in orbital energy due to tidal oscillations and mass loss, as $\Delta \epsilon_{\rm orb} \propto M_{\star}/R_{\star}$. This further accelerates orbital shrinkage in denser WDs, leading to earlier final disruption.

\subsection{Compositional Transition in Repeating TDEs}\label{sec:Compositional_Transition}
Each passage strips a spherical shell of depth $\Delta \simeq (1 - 0.5/\beta)R_{\star}$ from the outermost remaining layer of the WD \citep{2023ApJ...947...32C}. For $\beta = 0.7$, $\Delta/R_{\star} \sim 0.28$, which is deep enough to remove the H1 envelope of a He WD and to strip both the H1 and He4 envelopes as well as C12/O16 material from a CO WD (Figure~\ref{fig:Abundance profile}). For $\beta = 0.55$, $\Delta/R_{\star} \sim 0.1$, so only a thin layer is removed and the stripped debris from a CO WD is dominated by He4. These analytic estimates are consistent with our simulations. To approximate the bulk abundance removed in each passage of a repeating TDE, we assume that stripping leaves the normalized abundance profile of the surviving WD, $X(R/R_{\star})$, unchanged, and that the star expands to a new radius given by Equation~\ref{eqt:mass_radius_relationship} after each passage. Mapping the stripped depth $\Delta$ onto this profile yields the composition of the stripped mass in each encounter.

\begin{figure*}[htbp]
\centering
\gridline{\fig{ 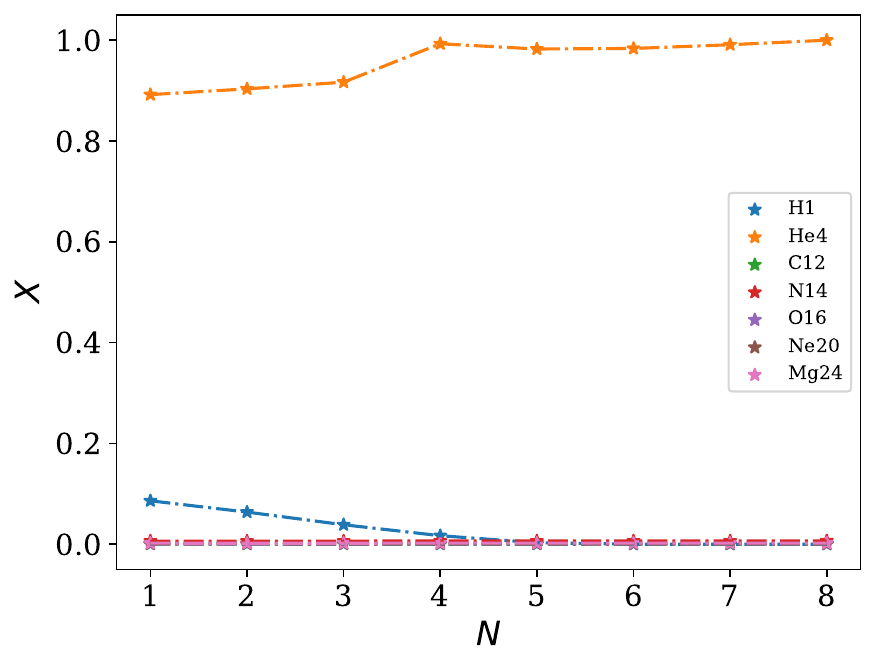}{0.48\textwidth}{}
          \fig{ 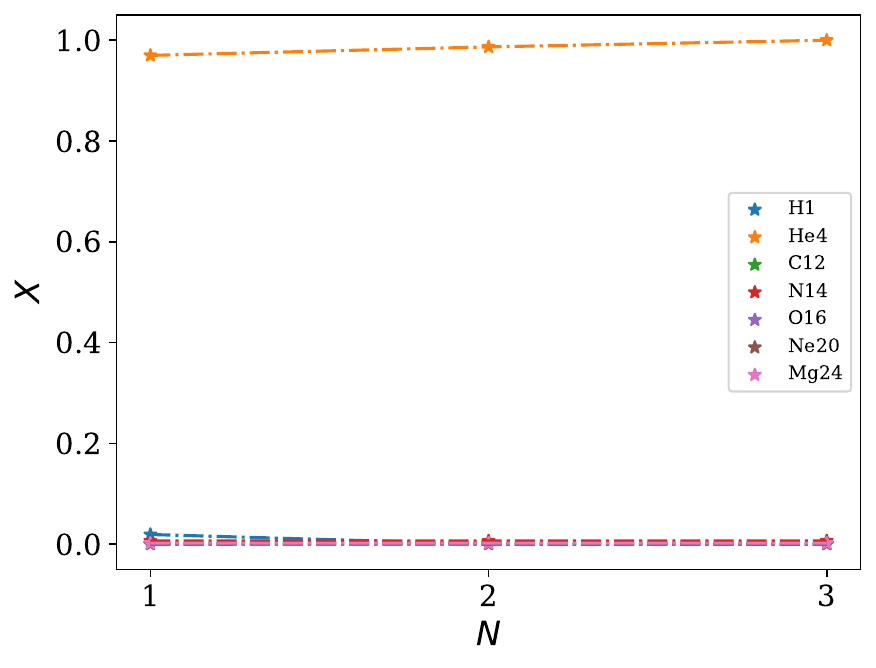}{0.48\textwidth}{}
         }
\gridline{\fig{ 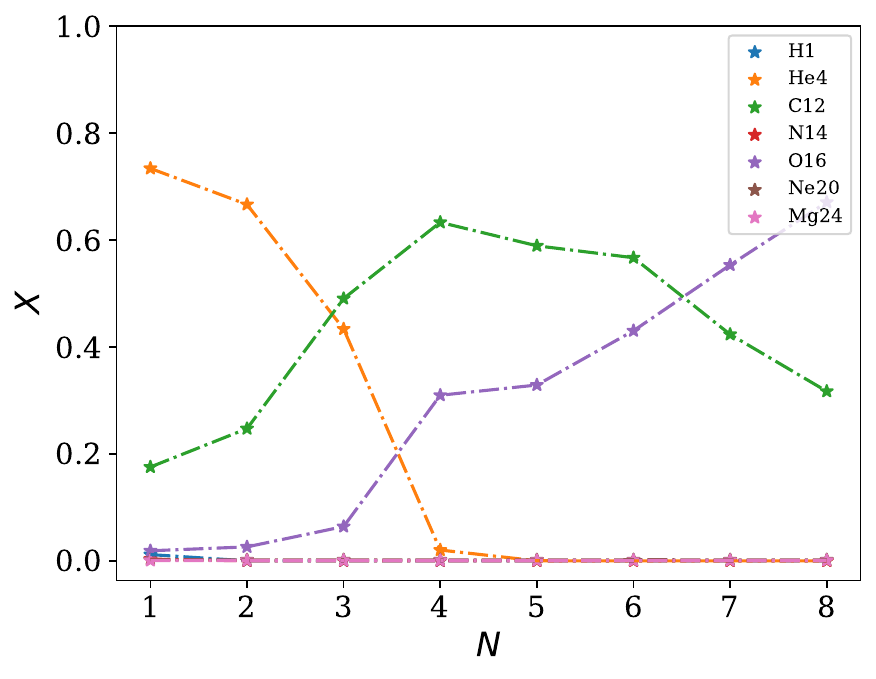}{0.48\textwidth}{}
          \fig{ 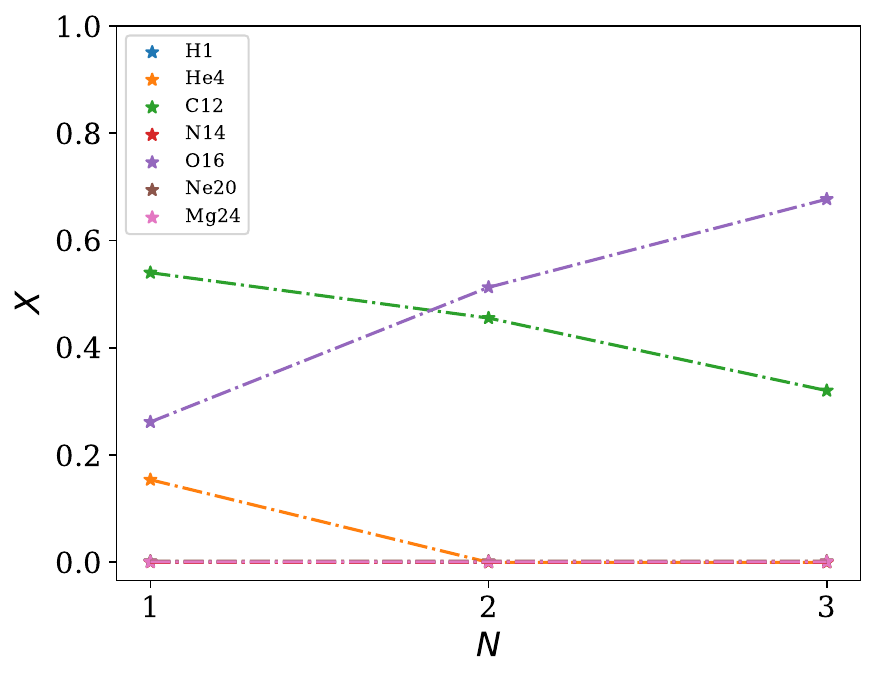}{0.48\textwidth}{}
         }
\caption{Compositional transition of the stripped mass for a $0.15\,M_{\odot}$ He WD (top) and a $0.6\,M_{\odot}$ CO WD (bottom) in repeating encounters with $\beta_0 = 0.55$ (left) and $\beta_0 = 0.7$ (right).\label{fig:stripped_mass_abundance}}
\end{figure*}

Figure~\ref{fig:stripped_mass_abundance} shows the predicted compositional evolution of the stripped mass for a $0.15\,M_{\odot}$ He WD and a $0.6\,M_{\odot}$ CO WD for $\beta_0 = 0.55$ and $\beta_0 = 0.7$. For $\beta_0 = 0.55$, both the He WD and the CO WD lose only a very thin envelope in the first several encounters (with $\sim 10\%$ H1 for the He WD and $\sim 50\%$ He4 for the CO WD). In the final few passages, the stripped composition becomes He-dominated for the He WD and CO-dominated for the CO WD.

For $\beta_0 = 0.7$, stripping from the He WD is He-dominated throughout. For the CO WD, the first encounter strips the outer CO core and the base of the helium envelope: $\sim 54\%$ of the stripped mass is C12, with $\sim 26\%$ O16 and $\sim 15\%$ He4. As shown in Figure~\ref{fig:element_specific_fallback}, the time-resolved fallback rate for this passage exhibits a He-to-C transition before $t_{\mathrm{peak}}$. Since the helium envelope is removed in the first encounter, the second passage strips directly into the CO core, producing a stripped mass of $\sim 46\%$ C12 and $\sim 51\%$ O16 with no hydrogen or helium. The fallback rate in this passage is therefore expected to show a C-to-O transition, with O16 dominating at peak. As the effective impact parameter rises and the WD becomes increasingly vulnerable, the third encounter results in full disruption, with stripped mass dominated by O16 ($\sim 68\%$) and a reduced C12 fraction ($\sim 32\%$). The C-to-O transition should thus become stronger and occur earlier in this final passage, reflecting deeper penetration into the oxygen-rich core.

The fallback composition is therefore progressively enriched in heavier elements with each passage, strengthening the associated spectral-line emission. \citet{2018ApJ...857..109G} suggested that such abundance variations in the fallback debris can be used to infer the progenitor's pre-disruption properties. Given the WD's stratified structure, spectroscopy of repeating TDEs offers a promising probe of WD interior composition and initial mass.

In this framework, for a WD of given mass and radius we can compute the stripped mass per passage and the total number of passages $N_{\max}$ the WD survives before disruption. Combined with the abundance profile, this yields the composition of the stripped mass in each encounter. 

\section{Discussion}\label{sec:discussion}
\subsection{Abundance transitions as observational diagnostics}

Our simulations reveal two related forms of compositional evolution. First, the fallback composition evolves within an individual encounter because debris originating from different layers of the WD occupies different ranges of orbital energy. For the CO-WD model, the grazing encounter with $\beta=0.55$ produces an early H-to-He transition, whereas the deeper encounters with $\beta=0.7$ and $0.8$ produce He-to-C transitions before peak fallback. The transition occurs earlier relative
to $t_{\rm peak}$ and becomes more pronounced as $\beta$ increases. Second, our semi-analytic extrapolation predicts a passage-to-passage
evolution in which later encounters sample progressively deeper and more
C/O-rich material.

The dimensionless transition time,
\begin{equation}
    \tau_{A\rightarrow B}
    \equiv \frac{t(\dot{M}_{A}=\dot{M}_{B})}{t_{\rm peak}},
\end{equation}
therefore provides a possible ``compositional clock'' for the depth of stripping. In our CO-WD calculations, the He-to-C transition moves from $\tau_{\rm He\rightarrow C}\simeq0.9$ at $\beta=0.7$ to $\simeq0.75$ at $\beta=0.8$. This mapping should not be regarded as a universal calibration of $\beta$, because the transition time also depends on the WD mass, envelope mass, cooling history, orbital eccentricity, and the numerical treatment of mixing. The more robust prediction is the ordering of the transitions: material from the outer H/He layers returns before increasingly C/O-rich debris from deeper layers.

Element-specific fallback rates do not map directly onto observed emission-line ratios. Line luminosities depend on the ionization state, density, optical depth, radiation field, geometry, and viewing angle of the line-forming gas \citep{2016ApJ...827....3R,2017ApJ...841..132L,
2018ApJ...857..109G}. Consequently, the fallback-rate ratios defined in
Section~\ref{sec:Composition dependence of Mass fallback Rate} quantify the available elemental mass supply rather than the strengths of
particular spectral lines. Photoionization and radiative-transfer calculations are required to determine whether, for example, an increase
in the C/He fallback ratio produces a detectable increase in the corresponding line ratio.

Subject to this caveat, the most promising signature is a correlated temporal sequence rather than the detection of a single element. A
grazing encounter should initially supply H-rich and subsequently He-rich material, whereas deeper encounters should show an increasing
contribution from carbon/oxygen before peak fallback. The proposed diagnostic should therefore be formulated in terms of time-dependent abundance ratios rather than the absence of a single line.

The observability of an abundance transition also depends on the response of the accretion flow. If circularization and inflow are rapid, the composition of the accreting gas may approximately track the instantaneous fallback composition \citep{chen_multi-messenger_2026}. 

Strong reprocessing in an optically thick disk wind may instead make the cumulative composition more relevant than the instantaneous fallback composition. High-cadence synthetic spectra are therefore needed before the transition times measured from our fallback curves can be compared quantitatively with observations.

\subsection{Implications for repeating disruptions}

Repeated stripping removes the WD from the outside inward, causing later flares to become progressively more enriched in heavier elements. For a CO WD, this suggests a qualitative sequence from H/He-rich debris in early grazing passages to C-rich and eventually more O-rich debris as the encounter deepens. Such passage-to-passage evolution would be more distinctive than a single composition measurement, because it would connect recurrent activity to the layered structure of the same donor.

Several assumptions limit the quantitative accuracy of these estimates. We approximate the removed material as an equivalent spherical shell, although tidal mass loss is intrinsically anisotropic. We also assume
that the remnant re-equilibrates according to the adopted mass--radius relation while retaining its normalized abundance profile. In reality, tidal heating, stellar oscillations, hydrodynamic mixing, and incomplete removal of the envelope at different latitudes may alter both the structure and composition of the next stripped layer. Direct multi-passage simulations are required to test whether the outside-in compositional sequence survives remnant mixing and repeated tidal heating.

If some QPEs are produced by repeated mass transfer from a WD, the passage-to-passage abundance evolution predicted here could provide an additional test of that
interpretation \citep{2020MNRAS.493L.120K,2022ApJ...930..122C,
2022ApJ...933..225W,2022A&A...661A..55Z,2023ApJ...947...32C}. Such a test will be challenging because many QPEs show little transient optical emission during individual eruptions, and optical spectra can be dominated by the host-galaxy nucleus. High-resolution soft-X-ray spectroscopy may instead constrain time dependent C, N, and O features in favorable
sources. Helium is intrinsically more difficult to identify in a hot, highly ionized X-ray-emitting plasma, so the absence of an observable
He feature would not by itself imply a He-poor donor. For nearby systems in the appropriate frequency range, a simultaneous low-frequency
gravitational-wave detection would provide an independent constraint on the orbital parameters and compact nature of the donor \citep{2014ApJ...794....9M}.

\subsection{Candidate events and the role of outflows and jets}

Several sources have been discussed as possible WD disruptions, including the ultraluminous X-ray (ULX) source in NGC~1399 \citep{irwin_evidence_2010}, a candidate near M86 \citep{jonker_discovery_2013}, and the fast X-ray transient EP250702a \citep{2026SciBu..71..538L}. The lack of hydrogen emission in this ULX candidates is suggestive to be the WD disruption evidence. EP250702a have also motivated WD--IMBH interpretations, although competing main-sequence-star disruption models remain
viable \citep{2026MNRAS.545f2019N,2026MNRAS.546ag005E,
2026MNRAS.547ag328G}. Luminosity and duration alone are therefore unlikely to identify a WD TDE uniquely. Time-dependent abundance information would provide a more direct connection to the internal structure of the disrupted object.

The high mass-supply rates expected in WD TDEs can produce super-Eddington accretion, powerful disk winds, and possibly relativistic jets \citep{2012ApJ...749...92K,2014ApJ...794....9M,
2015MNRAS.453..157P}. These components affect the abundance diagnostic in different ways. A beamed, nearly featureless jet continuum may dilute thermal emission and make composition measurements difficult, whereas a dense disk wind or reprocessing layer may carry the composition of the accumulated fallback material and provide a more favorable line-forming region. Simultaneous X-ray and UV spectroscopy, rather than photometry alone, will be needed to separate these possibilities.

\section{Conclusions}\label{sec:conclusion}

In this work, we investigated how the internal structure and chemical
stratification of WDs shape the time-dependent fallback composition in
partial tidal disruptions by an IMBH. We combined realistic
\texttt{MESA} WD models with three-dimensional hydrodynamic simulations
of individual encounters and used a semi-analytic framework to
extrapolate the evolution to repeated partial disruptions. Our main
findings are as follows:

\begin{enumerate}

\item For the $0.6\,M_{\odot}$ CO WD, the fallback debris exhibits a
$\beta$-dependent and time-resolved compositional transition that
reflects the layered structure of the WD. The grazing encounter with
$\beta=0.55$ produces an H-to-He transition at approximately
$0.3t_{\rm peak}$. The deeper encounters with $\beta=0.7$ and $0.8$
penetrate the CO core and produce He-to-C transitions at approximately
$0.9t_{\rm peak}$ and $0.75t_{\rm peak}$, respectively. Thus, the
He-to-C transition occurs earlier relative to peak fallback and the
carbon contribution becomes larger as the encounter deepens. By
contrast, the $0.15\,M_{\odot}$ He WD disrupted at $\beta=0.7$ remains
He-dominated throughout the resolved fallback evolution and shows no
clear H-to-He transition.

\item These abundance transitions provide a potential compositional
diagnostic of WD tidal disruptions.
The predicted evolution may manifest itself through time-dependent
abundance-sensitive features in the UV, optical, or soft-X-ray spectrum.
In particular, UV and optical spectroscopy may constrain the H and He
components in cooler reprocessing material, while high-resolution
soft-X-ray spectroscopy may probe partially ionized C, N, and O in a
disk atmosphere, outflow, or reprocessing layer.

\item Under the assumptions of our semi-analytic model, repeated
stripping removes the WD from the outside inward, so that the stripped
material becomes progressively enriched in heavier elements. For
$e_0=0.9$ and a near-threshold initial encounter with
$\beta_{\rm init}=0.5$, the seven WD models considered here survive
approximately $30$--$80$ passages before complete disruption. By
contrast, encounters beginning at $\beta_{\rm init}=0.7$ lead to
disruption within approximately three passages. In the
$\beta_{\rm init}=0.5$ calculations, the systems enter a rapid
mass-loss phase when the cumulative removed mass reaches approximately
$5\times10^{-2}M_0$.

\end{enumerate}

Our repeated-encounter predictions are anchored in single-passage hydrodynamic simulations. The semi-analytic extrapolation approximates the removed material as an equivalent spherical shell and assumes that the remnant re-equilibrates while retaining its normalized abundance profile. Direct multi-passage hydrodynamic simulations with realistic
WD structures are therefore needed to test the predicted passage-to-passage compositional evolution and the number of survivable
encounters. Combining such simulations with time-dependent photoionization and radiative-transfer calculations will be essential for converting the fallback-abundance transitions into quantitative spectral predictions.

Wide-field facilities such as \textit{Einstein Probe} for potentially TDEs, can provide rapid discovery and time-domain monitoring of
candidate fast X-ray transients. Testing the compositional predictions will additionally require rapid UV/optical spectroscopy and, for sufficiently bright and weakly absorbed sources,
high-resolution soft-X-ray observations capable of resolving C, N, and O features. Facilities such as ArgusSpec \citep{galliher_argusspec_2024}, rapid multiband photometry with TriCCS, and spectroscopy with KOOLS-IFU on the Seimei Telescope \citep{tsuzuki_development_2024}, as well as the New Robotic Telescope \citep{oria_4-m_2020}, may help characterize the minute-to-hour evolution predicted by our fallback calculations, provided that the events are sufficiently bright. Coordinated observations across these wavelengths offer the most promising route for testing whether time-dependent abundance transitions can identify WD disruptions and their associated IMBHs.

\begin{acknowledgments}
This work is supported by the National Natural Science Foundation of China (the NSFC Type C Youth Project 12503053). We acknowledge the support from the Hong Kong Research Grants Council (HKU 17305124, N\_HKU782/23). We also acknowledge the useful discussion with Bing Zhang. The FLASH code used for the simulations in this work was developed in part by the DOE NNSA and the DOE Office of Science-supported Flash Center for Computational Science at the University of Chicago and the University of Rochester. The authors would also like to thank the National Supercomputing Center in Guangzhou Center and Nansha Subcenter for providing high-performance computational resources. We thank the participants of the TDE FORUM (Full-process Orbital to Radiative Unified Modeling) online seminar series for their inspiring discussions.
\end{acknowledgments}

\counterwithin{figure}{section}
\twocolumngrid

\appendix
\section{Element-specific Fallback Rate Relative to Solar Abundance} \label{Appendix A}
\begin{figure*}[h!]
\centering
\gridline{\fig{ 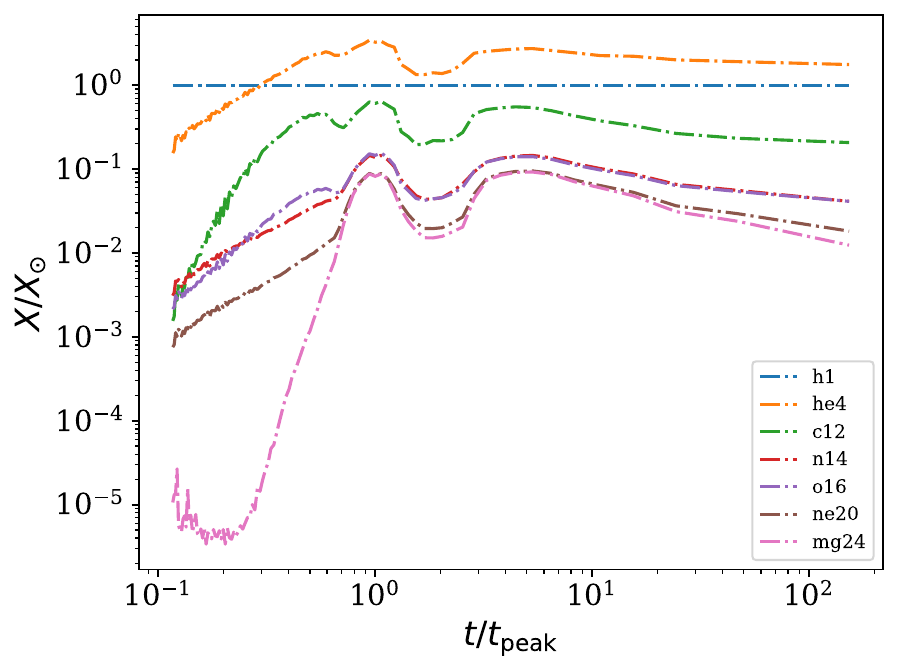}{0.33\textwidth}{}
          \fig{ 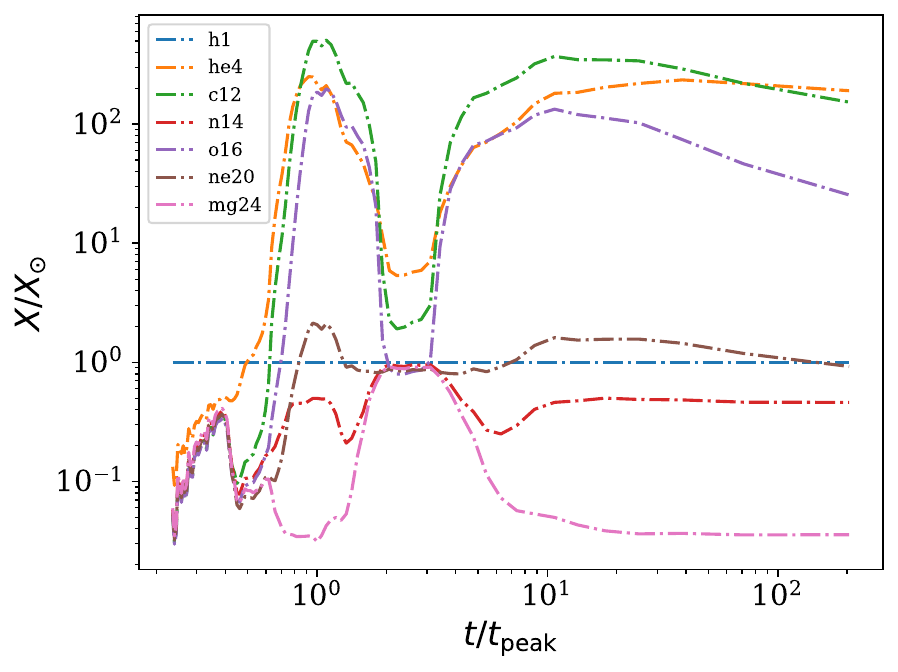}{0.33\textwidth}{}
          \fig{ 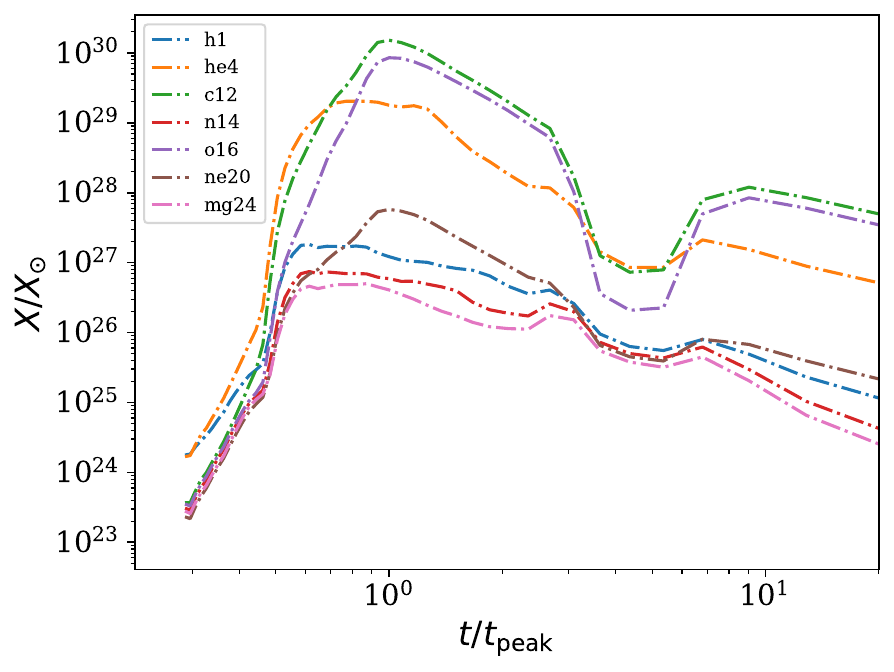}{0.33\textwidth}{}
         }
\caption{Fallback rate of different elements relative to the solar abundance for disruption of a $0.6\,M_{\odot}$ CO WD in the same encounters as Figure~\ref{fig:element_specific_fallback} (left to right: $\beta = 0.55, 0.7, 0.8$). \label{fig:abundance_evolution_rel_solar}}
\end{figure*}

We show the element-specific fallback rate of the CO WD in a tidal encounter relative to the solar abundance and the mass fallback rate. We adopt the solar abundance from \citet{2009ARA&A..47..481A}, and $X/X_{\odot}$ is given by this equation from \citet{2018ApJ...857..109G}, 
\begin{equation}
    \frac{X}{X_{\odot}} = \frac{\dot{M}_{\rm X}/\dot{M}_{\rm H}}{(X_{\rm X}/X_{\rm H})_{\odot}}, 
\end{equation}
where $(X_{\rm X}/X_{\rm H})_{\odot}$ is the abundance ratio relative to H in the solar abundance, and the numerator gives the abundance ratio relative to H in the fallback materials. Figure~\ref{fig:abundance_evolution_rel_solar} show the abundance relative to the solar abundance, which might provide insight into observations of TDEs when we do not know the type of the disrupted star. We can distinguish the disruption of WDs and Sun-like stars, and constrain the stellar types by comparing the elements observed in TDEs with the solar abundance. 

\section{Element-specific Fallback Rate Relative to Mass Fallback} \label{Appendix B}
\begin{figure*}[h]
\centering
\gridline{\fig{ 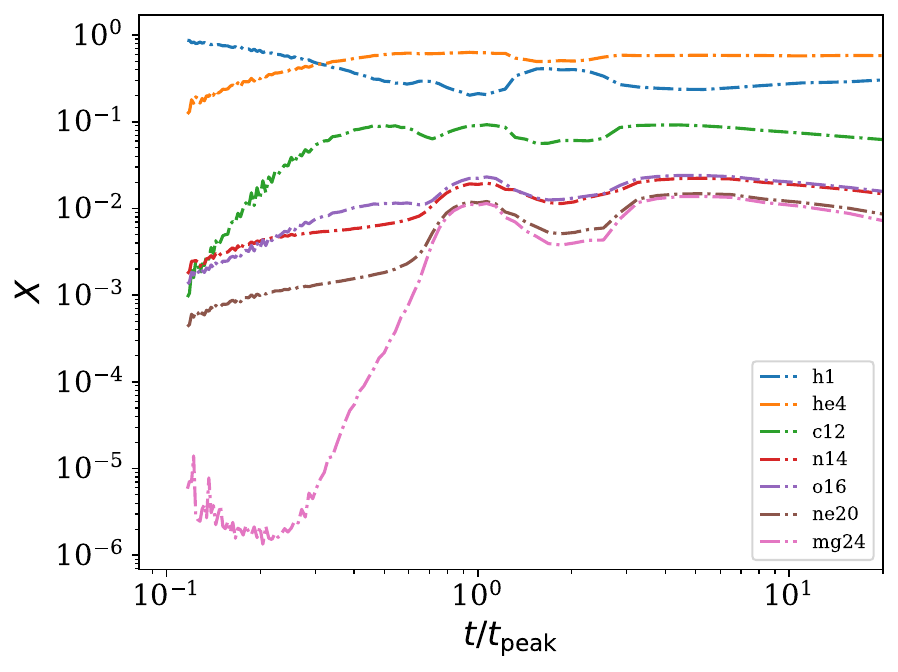}{0.33\textwidth}{}
          \fig{ 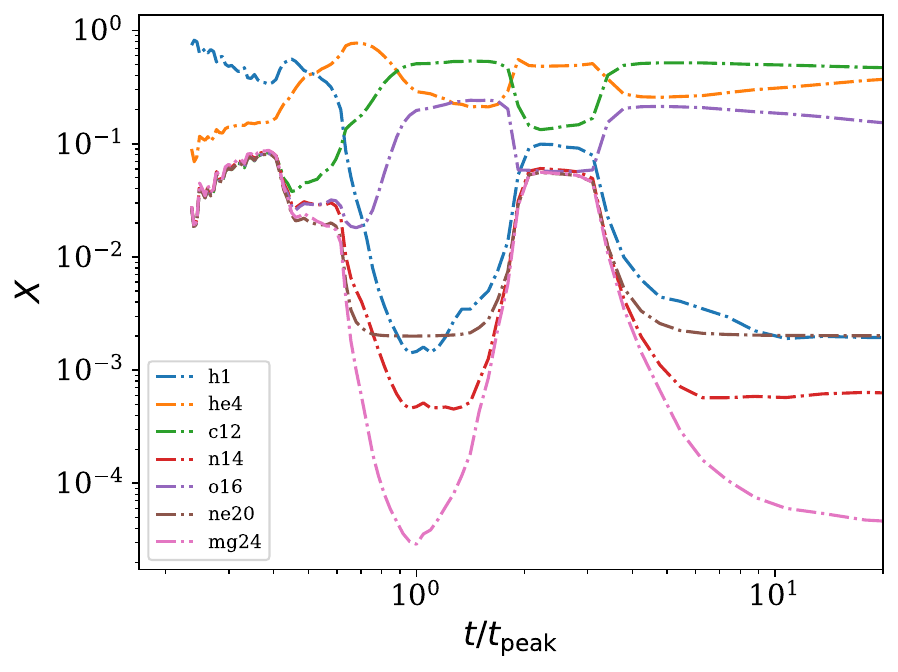}{0.33\textwidth}{}
          \fig{ 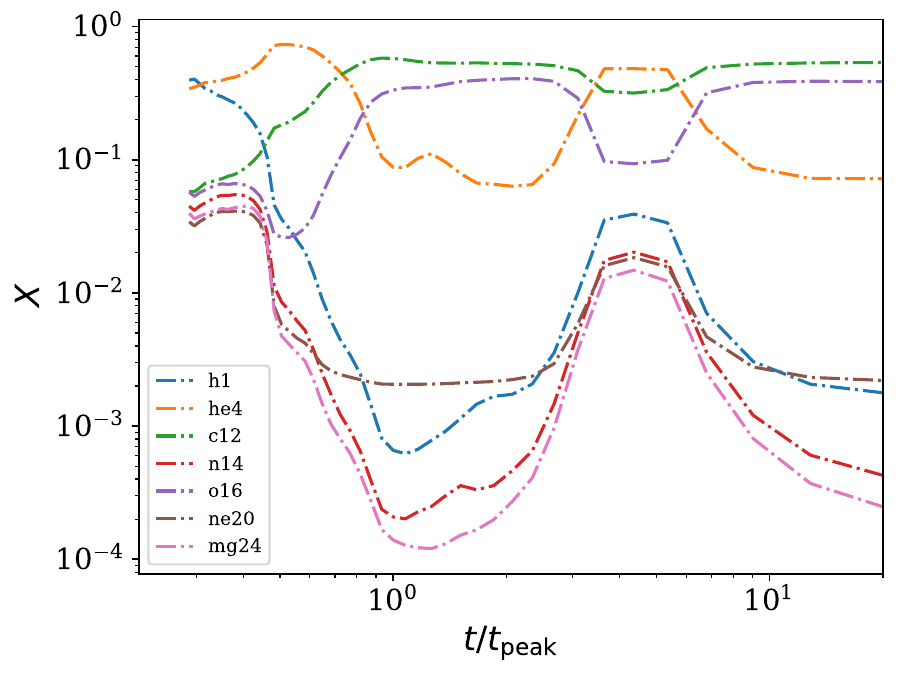}{0.33\textwidth}{}
         }
\caption{Same as Figure~\ref{fig:abundance_evolution}, but it is a log-log plot. (left to right: $\beta = 0.55, 0.7, 0.8$). \label{fig:abundance_evolution_rel_mass}}
\end{figure*}

Figure~\ref{fig:abundance_evolution_rel_mass} shows the abundance relative to the mass fallback rate; in other words, it illustrates the evolution of the abundance of fallback materials. This is the same as Figure~\ref{fig:abundance_evolution}, but on a log-log scale, allowing the evolution of less abundant elements to be seen more clearly. For strong encounters ($\beta = 0.7$ and $0.8$), the abundances of heavier elements (N14, Ne20, Mg24) increase when the remnant core returns to the pericentre. This is because the encounters penetrate the core, causing the heavier elements initially located inside the core to fall back. In contrast, for a weak encounter ($\beta = 0.55$), which strips only the envelope, the return of the remnant core does not bring heavier elements into the fallback material.

\bibliography{references}{}
\bibliographystyle{aasjournalv7}

\end{CJK*}
\end{document}